\documentclass{aa}

\usepackage{txfonts}
\usepackage{natbib}
\usepackage{graphicx}
\usepackage{mathtools}
\bibpunct{(}{)}{;}{a}{}{,} % to follow the A&A style
\usepackage{longtable}
\usepackage{xspace}
   \newcommand{\degree}{\ensuremath{^{\circ}}}

\begin{document}

\title{Possible ring material around centaur (2060) Chiron}

\author{J.L. Ortiz \inst{1} \and R. Duffard \inst{1} \and N. Pinilla-Alonso \inst{2} \and A. Alvarez-Candal \inst{3} \and P. Santos-Sanz \inst{1} \and  N. Morales \inst{1} \and E. Fern\'{a}ndez-Valenzuela \inst{1}  \and  J. Licandro \inst{4,5} \and  A. Campo Bagatin \inst{6}  \and  A. Thirouin \inst{7,1} }

\institute{$^1$ Instituto de Astrof\'{\i}sica  de Andaluc\'{\i}a, CSIC, Apt 3004, 18080, Granada, Spain. \\
$^2$ Department of Earth and Planetary Sciences, University of Tennessee, Knoxville, TN, 37996, United States \\
$^3$ Observat\'{o}rio Nacional, Rua General Jos\'{e} Cristino 77, Rio de Janeiro 20921-400, Brazil.  \\
%$^4$ Max Planck Institute for Solar System Research, Justus-von-Liebig-Weg 3, 37077, G\"{o}ttingen, Germany \\
%$^4$ Observatoire de Paris, LESIA, 92195 Meudon, France \\
%$^5$  Universit\'{e} Pierre et Marie Curie, 75252 Paris, France   \\
$^4$ Instituto de Astrof\'{i}sica de Canarias, c/V\'{i}a Lactea s/n, 38200 La Laguna, Tenerife, Spain \\
$^5$ Departamento de Astrof\'{i}sica, Universidad de La Laguna (ULL), 38205 La Laguna, Tenerife, Spain \\
$^6$ Departamento de F\'{i}sica, Ingenier\'{i}a de Sistemas y Teor\'{i}a de la Se\~nal, Universidad de Alicante, Carretera San Vicente del Raspeig S/N, 03690 San Vicente del Raspeig, Alicante, Spain  \\
$^7$ Lowell Observatory, 1400 W Mars Hill Rd, 86001, Flagstaff, Arizona, USA \\
}

\abstract{ We propose that several short duration events observed in past stellar occultations by Chiron were produced by rings material. Some similarities between these events and the characteristics of Chariklo’s rings could indicate common mechanisms around centaurs. From a reanalysis of the stellar occultation data in the literature we determined two possible orientations of the pole of Chiron's rings, with ecliptic coordinates $\lambda$=(352$\pm$10)$\degree$, $\beta$=(37$\pm$10)$\degree$ or $\lambda$=(144$\pm$10)$\degree$, $\beta$=(24$\pm$10)$\degree$ . The mean radius of the rings is (324$\pm$10)~km.  One can use the rotational lightcurve amplitude of Chiron at different epochs to distinguish between the two solutions for the pole. The two solutions imply a lower lightcurve amplitude in 2013 than in 1988, when the rotational lightcurve was first determined. We derived Chiron's rotational lightcurve in 2013 from observations at the 1.23~m CAHA telescope and indeed its amplitude is smaller than in 1988. We also present a rotational lightcurve in 2000 from images taken at CASLEO 2.15~m telescope that is consistent with our predictions. Out of the two poles the $\lambda$=(144$\pm$10)$\degree$, $\beta$=(24$\pm$10)$\degree$ solution provides a better match to a compilation of rotational lightcurve amplitudes from the literature and those presented here. We also show that using this preferred pole orientation, Chiron's long term brightness variations are compatible with a simple model that incorporates the changing brightness of the rings as the tilt angle with respect to the Earth changes with time.
%This means that Chiron's absolute magnitude depends largely on its ring system and we built tentative models to match the long term evolution of Chiron's brightness.
%This means that the brightness of Chiron's system and the tilt angle of the rings are related and implies that the brightness of the ring system must be comparable to that of the main body}.
%The minima of Chiron's brightness occurred near an aspect angle of 90$\degree$ when the rings were edge-on.
Also, the variability of the water ice band in Chiron's spectra in the literature can be explained to a large degree by an icy ring system whose tilt angle  changes with time and whose composition includes water ice, analogously to the case of Chariklo.
We present several possible formation scenarios for the rings from qualitative points of view and speculate on the reasons why rings might be common in centaurs. We speculate on whether the known bimodal color distribution of the centaurs could be due to centaurs with rings and centaurs without rings.  }

\keywords{Minor planets, asteroids: individual: Chiron, Planets and satellites: rings, Occultations, Techniques: imaging spectroscopy.}

    \maketitle
%
%________________________________________________________________

\section{Introduction}

(2060) Chiron (formerly 1977~UB) was the first of a new class of objects discovered in our Solar System, the so-called "centaurs". These objects orbit the Sun at distances between Jupiter and Neptune. Soon after the discovery of Chiron several decades ago, this centaur was reported to show cometary-like activity \citep{Meech1989}. It had several brightness outbursts and developed coma, like a comet \citep{Luu1990}. For this reason Chiron is listed as a Solar System object with dual designation: comet 95P/Chiron and minor planet (2060).

Because Chiron is a very interesting body, it has been extensively observed and studied with plenty of techniques in the past. From short-term photometry studies rotational lightcurves have been determined for Chiron at different epochs \citep{Bus1989,Marcialis1993,Lazzaro1997}, showing that it has an elongated shape that causes a double peaked lightcurve. A very high accuracy rotational period of 5.917813 hours has been determined \citep{Marcialis1993} and also some constraints on the shape of the body have been obtained \citep[e.g.][]{Groussin2004}.  Spectroscopically, Chiron has been shown to possess water ice features in the nIR, whose exact widths seem to vary according to different authors who observed in different epochs \citep{Foster1999,Luu2000,Romon-Martin2003}. These features even disappeared in 2001 \citep{Romon-Martin2003}. From thermal infrared observations and modeling, the equivalent diameter of Chiron has been shown to be around 180 km, although a considerable scatter exists in the literature \citep[e.g.][]{Groussin2004}.

After the recent and surprising discovery that another centaur, Chariklo, has two dense rings from stellar occultation observations \citep{BragaRibas2014}, we wondered whether other centaurs could have rings. Because the rings of Chariklo are so close to the main body  and subtend a tiny angle of around 0.08 arcsec, direct imaging detection of rings is challenging with current technology so stellar occultations are the best means to detect rings around small bodies. The only centaur, apart from Chariklo, for which stellar occultations have been successfully recorded in the past is Chiron, and curiously, the lightcurves of these occultations showed rapid secondary dimmings, that were interpreted as comet-like dust jets from the nucleus \citep{Elliot1995, Bus1996} in 1993 and 1994 occultations, and as symmetric jet-like features \citep{Ruprecht2013a, Ruprecht2013b} in a 2011 occultation. Jets seemed a natural explanation given that Chiron had shown cometary activity in the past. However, the width of the two features reported by \cite{Ruprecht2013a} were 3 km and 7 km respectively, separated by a distance of 10 to 14 km, which are nearly identical to the 6.6 km and 3.4 km widths and 9 km separation of the rings in Chariklo as reported by \cite{BragaRibas2014}. Our interpretation of the sharp events in the Chiron 2011 occultation lightcurves is that these features were caused by two rings with similar properties to those recently reported for Chariklo's rings by \citet{BragaRibas2014}. Here we present such an interpretation.

In section 2 of this paper we present a reanalysis of the occultation lightcurves available in the literature, from which we derive two possible pole orientations for the proposed ring system. In section 3, we show that the change in Chiron's water ice spectral features with time is consistent with the proposed rings.
%In section 2 of this paper we present a reanalysis of the occultation lightcurves available in the literature and determine two possible ring orientations. {\bf In subsequent sections we try to discriminate a preferred orientation for the rings from the two possibilities and show that one of them can explain a number of observations}. It can explain why the water ice spectral features on Chiron change with time (section 3).
In section 4 we show that the the ring system can account for a large fraction of  the overall long-term photometric behavior of Chiron.  In section 5 we present two new rotational lightcurves that we obtained from the CAHA 1.23~m telescope in 2013 and from the CASLEO 2.15~m telescope in 2000, and we show that the observed changes in the amplitude of the rotational lightcurves are consistent with the expectations based on the preferred orientation of the ring system. In section 6 we show more indications in favor of the ring system.

In the last section  we discuss our results and speculate on why Chiron, the second largest centaur after Chariklo \citep{Duffard2014a,Fornasier2013}  can also have rings and whether rings can be common in centaurs and other outer Solar System bodies. Besides, we present some scenarios of ring formation from qualitative points of view and present different ideas that can be explored in the future.

%__________________________________________________________________

\section{Reanalysis of stellar occultations}

Three stellar occultations by Chiron have been reported since its discovery. The first stellar occultation was on November 7th, 1993, at around 13:16 UT and the results were published in \citet{Bus1996}, where all the pertinent details are given.  We will briefly mention that the magnitude of the star that was occulted was 14 and that the telescopes used were all smaller than 0.6 m, except for one. Another stellar occultation was observed in 1994, on March 9th at around 23:29 UT, although the findings associated with this occultation were reported earlier than for the 1993 event \citep{Elliot1995}.  The magnitude of the occulted star was 11.95, brighter than the one that was occulted in 1993, and the telescopes used were larger than in 1993, but only one occultation chord was obtained. On 29 November 2011 at around 8:16 UT another occultation by Chiron was observed and reported by \citet{Ruprecht2013a} and \citet{Ruprecht2013b}. In this case the occulted star was around magnitude 14.8, but the two telescopes involved in the observations were large, of 2m and 3m in diameter and provided the best data set.

In the \citet{Elliot1995} paper,  the first paper on an occultation by Chiron, two clear short duration drops in the lightcurves were reported and a few other longer lasting shallow events were reported too. They were all attributed to dust jets like those of comets, which seemed a reasonable explanation because of the known cometary-like behavior of Chiron  and also because symmetrical features with respect to the central body were not observed. That paper perhaps influenced the conclusions of the subsequent \citet{Bus1996} paper and the interpretation by \cite{Ruprecht2013a} and \cite{Ruprecht2013b} of the occultation in 2011, which is the event that shows the most compelling evidence for a ring system so those results will be discussed first.

\subsection{Stellar occultation in 2011}

Two datasets were obtained in this occultation, one from the NASA 3m  InfraRed Telescope Facility (IRTF) and another one from the 2m Faulkes telescope. Both datasets are reported in \cite{Ruprecht2013a} and \cite{Ruprecht2013b}. The data obtained with the NASA-IRTF telescope on Mauna Kea clearly show not only the occultation of Chiron but two short symmetrical lightcurve drops prior and after the occultation of the main body (the symmetry is with respect to the center of Chiron's occultation). The drops were at $\sim$ 300 km east and west of the center of the body occultation. The other dataset, obtained from the 2~m Faulkes telescope at Haleakala on Maui, provided a very high time-resolution (0.2s), high signal-to-noise lightcurve,  higher than what is usually achieved in stellar occultations by outer solar system bodies. That lightcurve does not show the occultation of  Chiron's main body, but it shows two pairs of very sharp drops of the star light. All this is clearly illustrated in figures 2-2 and 2-3 of \citet{Ruprecht2013a}. The lightcurve from the Faulkes telescope is very similar to that caused by Chariklo's rings as observed from the Danish 1.5~m telescope \citep{BragaRibas2014}.  The width of the two features reported by \cite{Ruprecht2013a} were 3 km and 7 km respectively, separated by a distance of 10 to 14 km, which are nearly identical to the 6.6 km and 3.4 km widths and 9 km separation of the rings in Chariklo as reported by \cite{BragaRibas2014}. Also, the optical depths of the extinction features were very similar, 0.4 and 0.05 for Chariklo rings and 0.7 to 1 for Chiron. The similarity is remarkable.  The only exception to the similarities of the 2~m Faulkes telescope dataset on the occultation by Chiron compared to the 1.5~m Danish telescope dataset on the occultation by Chariklo is that the occultation of the primary body was not detected in the Faulkes data whereas it was in the 1.5~m Danish telescope lightcurve, but this difference has no implication regarding the rings.   A double ring system with a gap might possibly be related to the presence of an unknown shepherd satellite, as we hypothesize in the discussion section.

Using the timing information that can be obtained from the plots in Figure 2-3 of \citet{Ruprecht2013a} and using the 7~s offset that the authors applied to the Faulkes lightcurve (as explained in \citet{Ruprecht2013a} they applied that shift in order to align the occultation midpoint with that seen in the IRTF data) we can derive approximate times for the disappearance of the rings as seen from the Faulkes 2~m telescope in Haleakala and from the IRTF telescope at Mauna Kea.   A double ring system is not resolved in the IRTF dataset because of its low time resolution  (2 s) so the times derived from the plot correspond to the middle of the occultation time of the two  proposed rings. For the Faulkes data we have taken the mean time of the occultations of the two proposed rings so that we have something comparable with the data derived from the IRTF.

\begin{figure}
\includegraphics[width=9cm, angle=0]{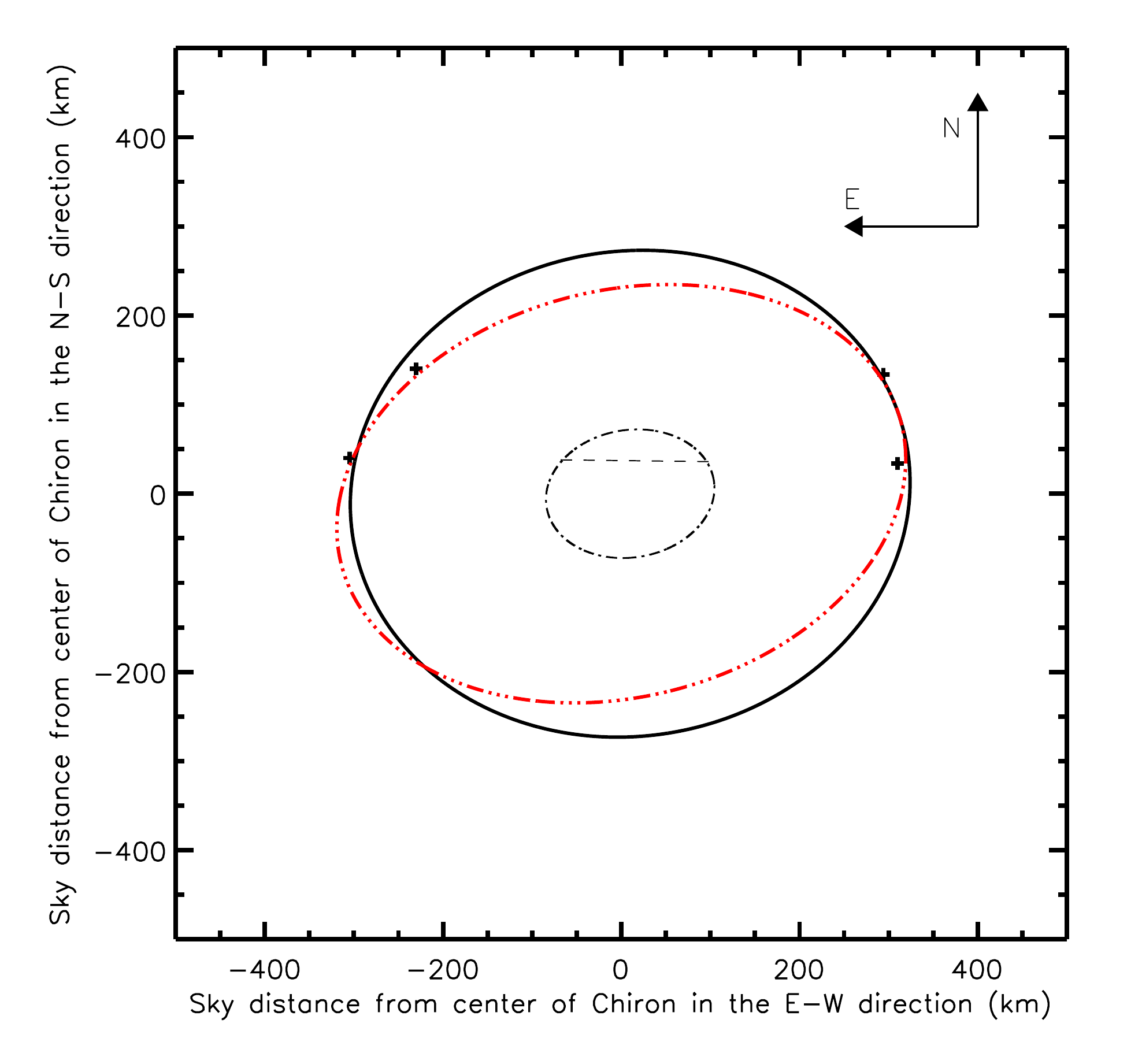}
\caption {\textit{Stellar occultation in 2011}: The plus symbols indicate the locations in the plane of the sky where the sharp events in the lightcurves of \citet{Ruprecht2013a} took place. The upper points correspond to events detected from the Faulkes 2m telescope in Haleakala, whereas the points below correspond to secondary events as seen from the IRTF 3m telescope at Mauna Kea. The ellipse in thick black line shows a possible fit to the secondary events within a range of possible fits that satisfy several constraints (see text). The semi-major axis is 313 km and the semi-minor axis is 272 km.  The ellipse in red line is an example of another possible fit. The semi-major axis of the ellipse is 324 km whereas the semi-minor axis is 228 km. The small dashed segment corresponds to the occultation chord of Chiron from IRTF. The smaller ellipse shows a tentative projected shape for Chiron that is compatible with the IRTF chord and is concentric with the ring system. The dimensions of this smaller ellipse are 190 km and 143 km. }
\label{fig1}
\end{figure}

Because we know the coordinates of the two observing sites we can project the secondary events in the plane of the sky. We can also plot the chord corresponding to the body detection observed from the IRTF (but not detected from the 2~m Faulkes telescope). An infinite set of ellipses corresponding to a ring system can be fitted to the four secondary mean events in the sky plane in Figure~\ref{fig1}. This is because an ellipse has 5 parameters and we only have four datapoints.

 But we can add a further constraint because we know that rings are made of material in circular orbits around the main body, so the center of the body must coincide with the center of the ellipse of the rings. Also, we know that Chiron's shape cannot be spherical \citep[because Chiron shows a double-peaked rotational lightcurve e.g.][]{Bus1989}, hence we can impose the constraint that the center of the ellipse be below the Mauna Kea chord and be slightly displaced with respect to the center of the chord.
%But because the rings must be concentric with Chiron and we also know that Chiron is not spherical (because Chiron shows a double-peaked rotational lightcurve e.g. \citet{Bus1989}), we can impose a further constraint that the center of the ellipse be below the Mauna Kea chord and be slightly displaced with respect to the center of the chord.
This results in a set of valid solutions for a mean ellipse representing the ring system. Valid fits are obtained for ellipses with a range of position angles from approximately 11$\degree$ to 27$\degree$. The range of major axis valid solutions is 314~km to 334~km (these values correspond to the distance of Chiron's center to the ring system, not to any of the two individual components). Hence 324~km is the average rings diameter. The valid aspect angle  or tilt angle of the ring system is approximately 31$\degree$ to 45$\degree$, under the assumption that the ring system is circular and we see it as an ellipse due to projection. Also the supplementary angles are possible, which implies that the range of angles from 135$\degree$ to 149$\degree$ is also valid.  By definition, the aspect angle is the angle that the rotation axis of a body (or rings in this case) make with respect to the direction defined by the observer and the body.
%is the Earth-body-pole of the body angle (this is the definition e.g in Tegler et al. 2005)
Using those angles we can determine approximate coordinates for the pole of the rings (which should coincide with the pole of Chiron if we assume that the rings are equatorial) with an uncertainty of around 10$\degree$. Two valid sets of pole direction solutions are possible because the tilt angle of the ring system has two valid sets of angles, as mentioned before.  The two valid solutions have been derived with an uncertainty of $\sim$10$\degree$. The first solution for the direction of the pole has ecliptic coordinates $\lambda$=(352$\pm$10)$\degree$, $\beta$=(37$\pm$10)$\degree$ and the other solution has coordinates $\lambda$=(144$\pm$10)$\degree$, $\beta$=(24$\pm$10)$\degree$.

The dimensions of the smaller ellipse shown in figure 1, are 190 km and 143 km for the major and minor axes respectively. This is a tentative ellipse that is compatible with the IRTF chord and is a reasonable shape for Chiron's limb. The uncertainties in those values are high compared to the the typical uncertainties obtained from stellar occultations with high time resolution and high time accuracy. The reason is because only one chord on the main body was obtained and as stated above, the center of the ellipse constraint is not very tight, so families of ellipses are compatible with the sharp secondary events. The equivalent diameter of the 190 x 143 km ellipse is somewhat smaller than the 218 $\pm$ 20 km equivalent diameter determined from Herschel Space Observatory measurements \citep{Fornasier2013}, but thermal emission of the rings and dust around Chiron might have contributed and was not separately accounted for in \cite{Fornasier2013}. Besides, there are other equivalent diameter estimations from thermal modeling in the literature that indicate diameters considerably smaller than 218 km for Chiron.  We have compiled different values from the literature: \cite{Fernandez2002} determined a diameter of 148 $\pm$ 8 km, \cite{Groussin2004} found 142 $\pm$ 10 km, \cite{Lebofsky1984} found 180 $\pm$ 40/50 km and \cite{Altenhoff1995} found 198 $\pm$ 20 km. On the other hand \citet{Campins1994} found values of 148 $\pm$ 22 km in 1991, 176 $\pm$ 20 and 208 $\pm$ 20 km in 1993, and, 188 $\pm$ 12 and 182 $\pm$ 26 km in 1994. All these results are lower than the Herschel value of 218 km and the scatter may indicate that Chiron was observed at different rotational phases (and that Chiron could be a rather elongated body), or maybe the rings contributions or the contributions from dust could be playing a role. But also differences in the thermal modeling by different authors and even mistakes or problems with the observations might be playing a role in giving rise to a considerable scatter.

We must note that the IRTF lightcurve also shows another brightness dip at around 8.262 UT of similar depth to one of the secondary symmetrical events of the rings. Hence, this event appears to be real, not noise. If it is real, it might be due to a small satellite or due to dust or even to an incomplete ring (an arc) because it does not have a symmetrical counterpart after the occultation of the main body. It is also curious that the number of dimming events in the IRTF and Faulkes lightcurves is apparently larger than the number of brightening events. In a pure noise distribution we should expect the number of dimming events to be nearly identical to the number of brightening events of the same magnitude, but this does not seem to be the case in these observations. This might be a hint on the existence of plenty of debris in different orbits around Chiron. However one should be cautious because large seeing changes can sometimes produce dimmings in synthetic aperture photometry.

\subsection{Stellar occultations in 1994 and 1993}

% we want to show that the ring system is not incompatible with the occultation features observed in 1993

\begin{figure*}
\includegraphics[width=18cm, angle=0]{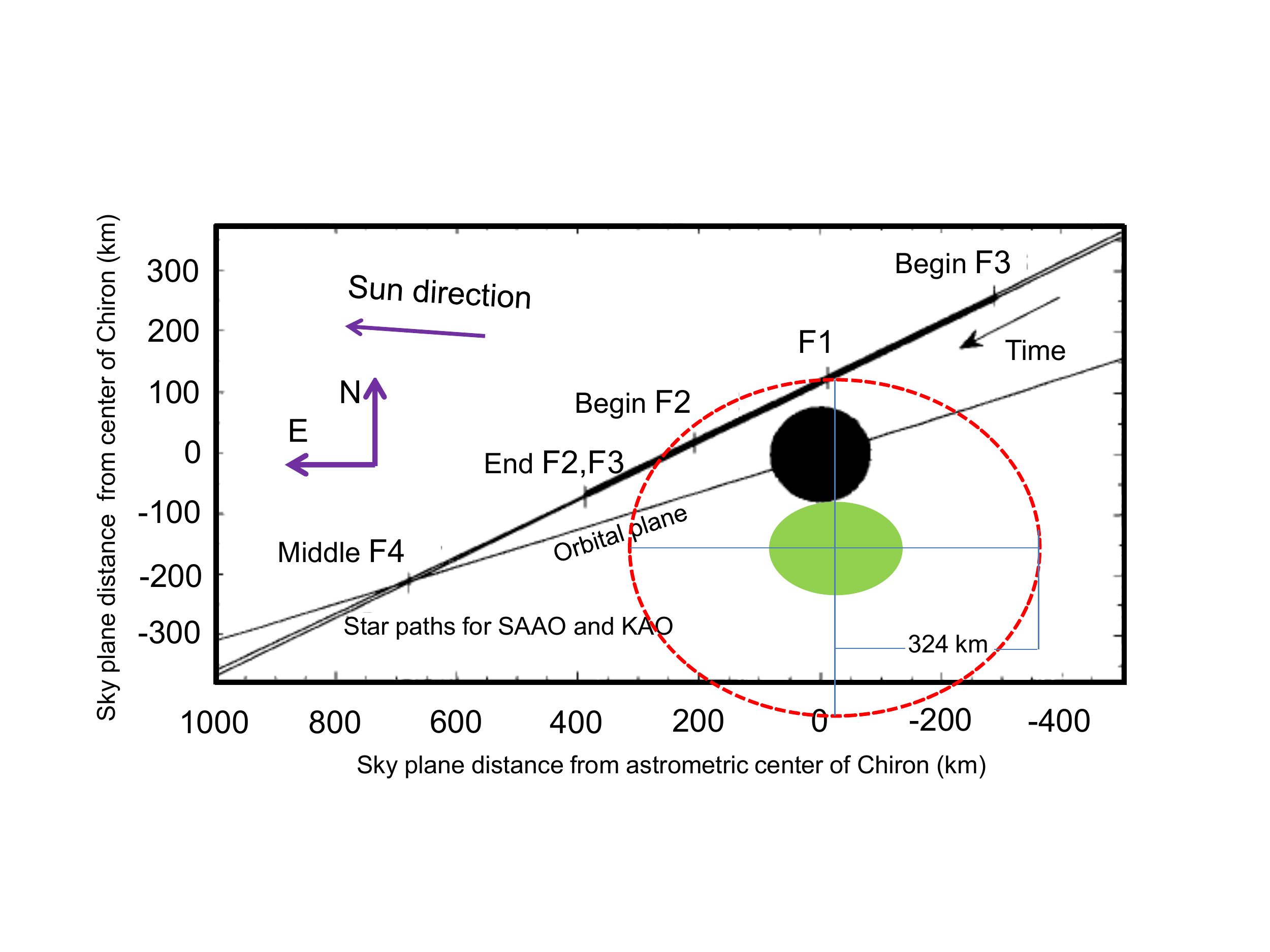}
\caption {\textit{Stellar occultation in 1994}: Adapted figure 2 of \citet{Elliot1995} showing the plane of the sky centered on Chiron (whose center was determined astrometrically, not from the occultation). The location of relevant lightcurve features are indicated with labels F1,F2,F3,F4. See \citet{Elliot1995} for the description of the features and to get relevant data regarding the occultation.  The black incomplete circle corresponds to the position where a spherical Chiron should have been with respect to the star paths for SAAO and KAO according to astrometric measurements. The orbit plane of Chiron is shown for reference. The red ellipse is the projected shape of a ring system whose pole was determined in 2011. The corresponding position angle of the ellipse in 1994 is 1 degree and the aspect angle is 30 degrees or 150 degrees for the two pole solutions. The intersections of this ellipse with the star paths from the two observing sites indicates where sharp occultation events would be detectable. As can be seen, the ring would clearly cause the F1 feature and also the F2 feature (whose maximum sharpness was not at the beginning of the label F2 in this figure, but slightly later, see text). The filled ellipse corresponds to a plausible position and shape of Chiron.}
\label{fig2}
\end{figure*}

 The next step is to compare the two proposed ring systems (the two orientations) to previous occultation data in order to see if they are compatible with the observations. Although the 1994 data are poorer in quality than in 2011 and some details cannot be completely determined, it makes sense trying to check whether the rings are incompatible or not with the 1994 observations. The two ring configurations resulting from the 2011 occultation  are compatible with the sharp lightcurve feature reported in \citet{Elliot1995}, with label \textit{F1} and possibly the feature with label \textit{F2}. This is shown schematically in Figure~\ref{fig2}, which is an adapted version of Figure 2 of \citet{Elliot1995} in which we incorporate an approximate ring position compatible with the observations and the position of the main body. In our version we need to shift the main body to the southwest compared to the \citet{Elliot1995} plot. This is not a problem because the position of the main body was not determined by \citet{Elliot1995} based on the occultation itself, but based on astrometry, which is typically accurate to the $\sim$20~mas level in optimal situations. So a displacement of around $\sim$10~mas to the southwest, which we require, is perfectly conceivable.  The optical depth of the F1 feature is 0.92 according \citet{Elliot1995} and the optical depths of the 2011 features are around 0.7 to 1, according to \citet{Ruprecht2013a}. These are very similar values in terms of extinction and are also of very similar widths (5 to 9 km and 3 to 7 km respectively, according the respective authors). Note that feature \textit{F2} in \citet{Elliot1995} is broader than the extremely sharp \textit{F1} feature, with a width of 74 km, but the deeper part of \textit{F2}, which is 3 to 4 times smaller, can be consistent with a ring feature. Maybe this feature is broader than \textit{F1} because it takes place in a part of the rings where there are fewer particles so a drop of light as sharp as in \textit{F1} should not be expected. It is possible that the ring system is not homogeneous azimuthally. \citet{Elliot1995} also mention the existence of two other features labeled \textit{F3} and \textit{F4} in their occultation lightcurves, but those features are very broad and shallow and they are hardly significant above the noise; if real, they are not associated to the ring system for which much sharper brightness drops are needed. They could be associated with dust or debris around Chiron.  \citet{Elliot1995} proposed a gravitationally bound coma to explain these very broad features.

%\subsection{Stellar occultation in 1993}

\begin{figure}
\includegraphics[width=9cm, angle=0]{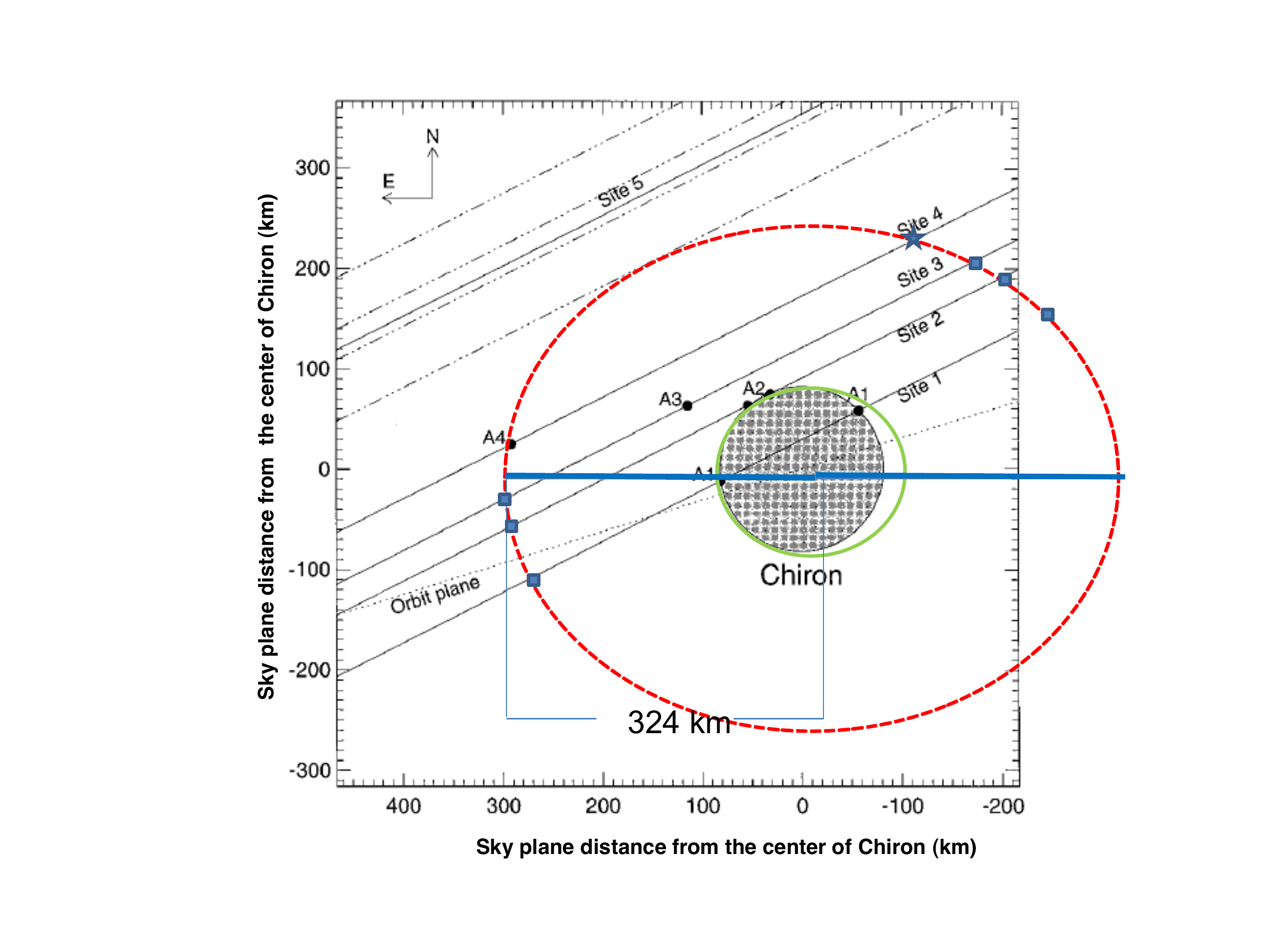}
\caption {\textit{Stellar occultation in 1993}: Adapted figure 5 of \citet{Bus1996} in which the sky plane for the region surrounding Chiron is shown. The ellipse in dashed line shows the ring of Chiron that would cause the A4 feature  and is compatible with the orientation derived from the two pole solutions derived from the 2011 occultation. The position angle is 1 degree and the aspect angle is 30 degrees or 150 degrees. The star symbol indicates another intersection of the ring with the star path for site 4, where another secondary event should have been detected from site 4 (see text), and the small square symbols show locations where secondary events should have been detected if the observations at the particular observing sites obtained data with high signal to noise ratio, which was not the case. The disk of Chiron is shown based on a circular fit to the occultation chords, but in green we show an alternative disc of Chiron. See main text.}
\label{fig3}
\end{figure}

Concerning the 1993 occultation, the ring configuration can be compatible with the sharp feature labeled \textit{A4}  Figure~\ref{fig3} in \citet{Bus1996}. In fact, the \textit{A4} feature is consistent with the \textit{F2} feature of \citet{Elliot1995}.  This A4 feature is <24 km in width, according to \citet{Bus1996}. One must also note that another ring extinction feature should have been detected in site 4 (star symbol), but the feature, which should have been analogous to the \textit{F1} feature in \citet{Elliot1995} was perhaps not detected because of the longer integration time (1s) used at site 4, in comparison to the observations in \citet{Elliot1995}. Hence, the very sharp and very brief \textit{F1} would not be  detected with the much poorer signal to noise ratio and poorer time resolution of \citet{Bus1996}. Note that the star that was occulted in 1994 was brighter than m$_V$=11.9 mag, whereas the star occulted in the 1993 event was around 2 magnitudes fainter. In summary, detecting sharp features in 1993 would have been difficult. An alternative explanation for the lack of a clear detection of a counterpart feature of \textit{F1} could be that the ring system is not homogeneous and can have gaps or regions with fewer particles (which rotate with the orbital period of the particles in the rings). If the ring system were homogeneous, other lightcurve features associated with the rings should have been recorded from sites 1, 2 and 3 (indicated with square symbols in fig. 3), but we must note that the telescopes employed at those sites were all smaller than the telescope used for the detection of the \textit{A4} feature, hence the poor signal-to-noise ratio and the longer integration times used in those sites would have prevented the detection of the ring features. This also happened in the occultation by the rings of Chariklo where several stations used too long integrations and too small telescopes to clearly detect the rings occultations \citep{BragaRibas2014}. Feature \textit{A3} in \citet{Bus1996} is dubious because the brightness drop does not seem to be below the 1-sigma noise level, and does not seem real at all compared with the stronger \textit{A4} feature. But even if \textit{A3} were real, it might be related to dust or debris around Chiron, not to the ring system.
In figure 3 we show in green line an alternative disc for Chiron to the circular fit proposed in \citet{Bus1996}. The shape of Chiron is known to be non-spherical hence the projection should not be a circle. The shown elliptical limb for Chiron is concentric with the ring's ellipse. The proposed ellipse is just for illustration. It is not a real fit to the chords, but it illustrates that non-spherical shape solutions are indeed possible to explain the observed chords. Because Chiron is small compared to a dwarf planet its shape is very likely somewhat irregular, so slight deviations from a pure ellipsoid are also expected. Hence, perfect fits to the chords with no residuals using a pure ellipse are unlikely.

 Combining the information from both occultations in 1993 and 1994, we should point out that for a feature at approximately the same distance from the center in both occultations statistical likelihood is higher for a fixed ring system than a jet. Hence our view is that at least some of the extinction features observed in 1993, 1994 and 2011 are compatible with the ring system and not compatible for a jet. We should keep in mind that Chiron was not reported to be active in those years, as no outbursts or coma was reported, so it is difficult to envision how a jet could have formed. Also, the dust size distribution derived in the 1994 occultation by \citet{Elliot1995} is very different to the typical size distribution in cometary jets observed from the ground and in particular to the size distribution of comet Halley observed by the Giotto spacecraft \citep{Lamy1987}. On the other hand the very narrow ejection angle needed for the jet has never been observed in comets. These and other points raised in section 6 allow us to reject the jet interpretation for the narrow and dense extinction features. A moderate range of possible position angles and aspect angles of the ring system in the 1993 and 1994 occultations are compatible with the constraints from the rings size and from the requirement that the rings must be concentric with the main body. In theory, that range of angles can be used to refine the direction of the pole derived from the 2011 occultation. However, no improvement has been obtained.

%Two valid solutions have been derived with an uncertainty of $\sim$10$\degree$. The first solution for the direction of the pole has ecliptic coordinates $\lambda$=(352$\pm$10)$\degree$, $\beta$=(37$\pm$10)$\degree$ and the other solution has coordinates $\lambda$=(144$\pm$10)$\degree$, $\beta$=(24$\pm$10)$\degree$.
%The preferred solution is the first solution reported here. The other solution can be rejected because it cannot explain the lack of a water ice spectral feature in 2001, whereas the $\lambda$=(352$\pm$2)$\degree$, $\beta$=(37$\pm$2)$\degree$ pole direction can explain not only that, but also the amplitudes of the rotational lightcurves. All this is described in subsequent paragraphs.

\section{Constraints from published spectra}

The spectra of Chiron reported in the literature show variable water ice spectral features at 2.02 $\mu$m and 1.5$\mu$m  \citep{Foster1999,Luu2000,Romon-Martin2003}. The water ice feature even disappeared in spectra taken in 2001, as reported by \citet{Romon-Martin2003} using spectra from the 8 m VLT, whereas the water ice feature could be detected only a few years earlier (in 1997) by \citet{Foster1999} despite their using a smaller telescope. We can explain the water ice spectral feature and even its disappearance with a ring system.

\begin{figure*}
\includegraphics[width=9cm, angle=0]{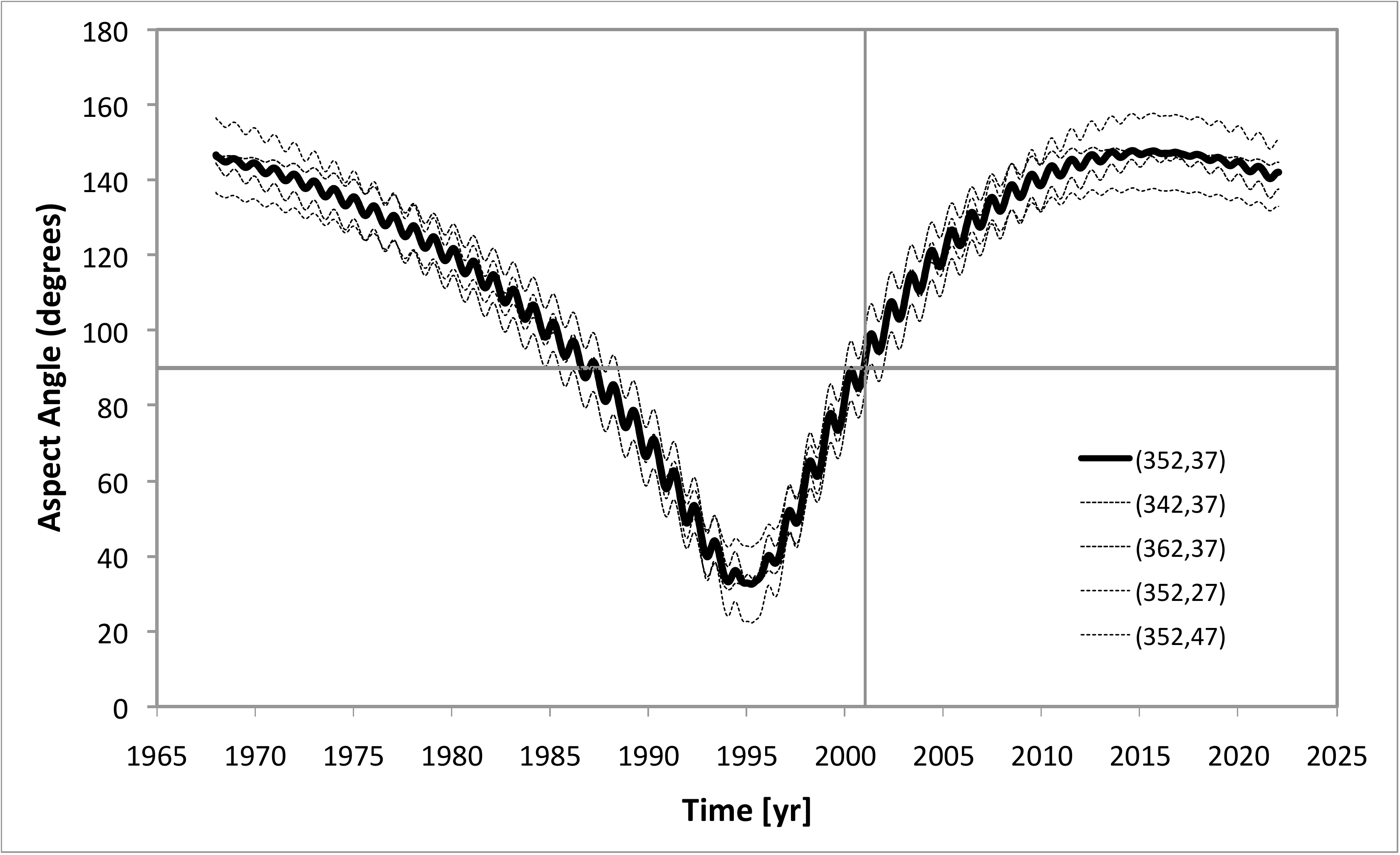}
\includegraphics[width=9cm, angle=0]{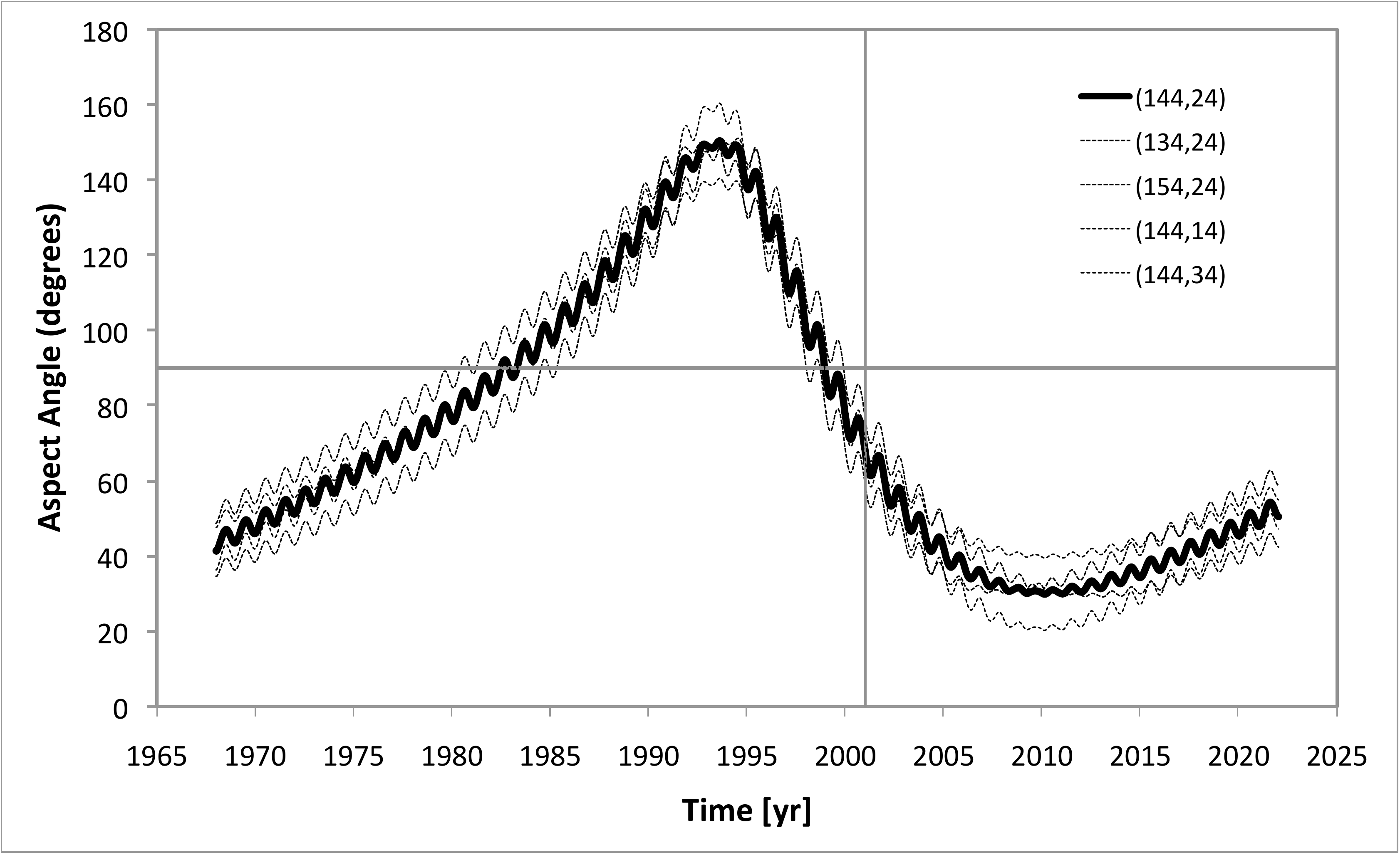}
\caption {\textit{Aspect angle versus time}: Left panel: The continuous line shows the aspect angle of Chiron's rings as a function of time for the nominal pole direction that we obtained from the analysis of the occultations of the rings ($\lambda$= 352 $\pm$10 $\degree$, $\beta$= 37 deg $\pm$ 10 $\degree$).  The rest of the lines correspond to solutions within the uncertainties of the nominal pole. The vertical and horizontal lines are shown to highlight that in 2001 the aspect angle of the rings was 90 degrees. The right panel shows the aspect angle as a function of time for the other pole direction ($\lambda$=(144$\pm$10)$\degree$, $\beta$=(24$\pm$10)$\degree$).}
\label{fig4}
\end{figure*}

Because we know that the rings of Saturn show prominent spectroscopic water ice features, we think that the water ice in Chiron is in the proposed rings, not on Chiron's surface.  Under the assumption that the water ice is only in the rings, the mechanism that can explain the disappearance of the water ice feature is the change of the aspect angle of the rings. This aspect angle is very close to 90$\degree$ in 2001 (see Figure~\ref{fig4}) if we use the first orientation of the ring system ($\lambda$=(352$\pm$10)$\degree$, $\beta$=(37$\pm$10)$\degree$), and around 70$\degree$ for the second ring pole orientation ($\lambda$=(144$\pm$10)$\degree$, $\beta$=(24$\pm$10)$\degree$). For both ring orientations the large decrease in the cross section of the rings can explain the disappearance of the water ice feature in 2001. The changing behavior of the water ice absorption features in Chariklo has been recently shown to be a result of the changing aspect angle of Chariklo's rings \citep{Duffard2014b}. Thus, in Chiron we would be seeing the same effect.

 According to \citet{Groussin2004} the composition that best matches both the observed reflectance spectra and albedo of Chiron corresponds to an aereal mixture of 30\% of water ice plus 70\% of refractory material and a common grain size of 10 $\mu$m. In other words, 30\% of the area of Chiron would be covered with ice and 70\% of the area would be covered with dark material. Because the geometric albedo determined by \citet{Groussin2004} was 0.11 and because the typical albedo of the dark material ($p_d$) in outer solar system bodies is around 0.05 to 0.08, the albedo of the water ice particles ($p_w$) can be obtained from $0.11=0.7 p_d + 0.3 p_w$. Hence $p_w$=0.27 if we use $p_d$=0.05 and $p_w$=0.19 if we use $p_d$=0.08. If we assume that the water ice particles are all in Chiron's rings, the work by \citet{Groussin2004} implies that the rings must be at least 30\% of the total area of Chiron's system. But we think that the area of the rings is even slightly larger than 30\% of the total area. The area of Chiron's main body is $\pi r^2$ with $r$=90 km (an intermediate size of all the values mentioned in the previous section) whereas that of the rings is approximately $2\pi r_{ring} w_p$ where $r_{ring}$ is the radius of the rings (324 km), $w_p$ is the combined width of the two rings seen from the observer (10 km, from the 2011 occultation). This means that the area of the rings  can be around 50\% of the total Chiron system. With the new percentages the albedo of Chiron's system is $0.11=0.5 \times 0.05 + 0.5 \times p_w$ so we can derive the albedo of the ring particles $p_w$. The new value is $p_w$=0.17 if we use $p_d$=0.05 and $p_w$=0.14 if we use $p_d$=0.08. Thus we think that the rings geometric albedo is in the 0.14 to 0.17 range based on the spectroscopy constraints, but could be up to 0.27 if the rings fractional area is around 30\% of the total system. This range of albedo values will be used in the next section to model Chiron's historical brightness evolution.

%As done for Chariklo \citep{Duffard2014b} one can model the existing spectra of Chiron in the literature at different epochs using the rings as the only source of water ice to see.
%{\bf We can use the widths of the two different ring components of Chiron (3 km and 7 km) to put constraints on the albedo of the rings in a similar way as done for Chariklo. We get equivalent geometric albedos for the rings in the range of 0.15 to 0.19 at 550 nm to explain the water ice spectral feature.}

 Because the 2001 spectra were taken while Chiron had experienced a clear activity outburst, it might be argued that the dust released in the outburst would dominate the reflected light and hence this could hide the water ice spectral feature. However, we have simulated this effect using the measured brightness increase of Chiron in 2001 and assuming that all this increase went to spectrally featureless dust. This means changing the areal mix in the spectroscopy models to around 85\% of spectrally featureless material. Even though the water ice spectral feature gets somewhat diluted in the simulations  with 85\% of spectrally neutral dust \citep[see fig. 8 of][]{Groussin2004}, the water ice band is still detectable, so the outburst cannot completely hide the water ice feature. Hence another mechanism must be in action or combined to the outburst to hide the water ice spectral feature.  Maybe an instrumental problem is the explanation, but the change of the aspect angle of the rings mentioned in the first paragraph of this section is the easiest explanation.

%The mechanism is the change of aspect angle of the rings, which is very close to 90 $\degree$ in 2001 (see Figure~\ref{fig4}) if we use the first ring orientation ($\lambda$=(352$\pm$10)$\degree$, $\beta$=(37$\pm$10)$\degree$), and around 70 $\degree$ for the second ring pole orientation ($\lambda$=(144$\pm$10)$\degree$, $\beta$=(24$\pm$10)$\degree$). For both ring orientations the decrease in the cross section of the rings together with the brightness outburst in 2001 is enough to make the water ice feature undetectable in 2001.

\section{Constraints from long term photometry}

\begin{figure}
\includegraphics[width=9cm, angle=0]{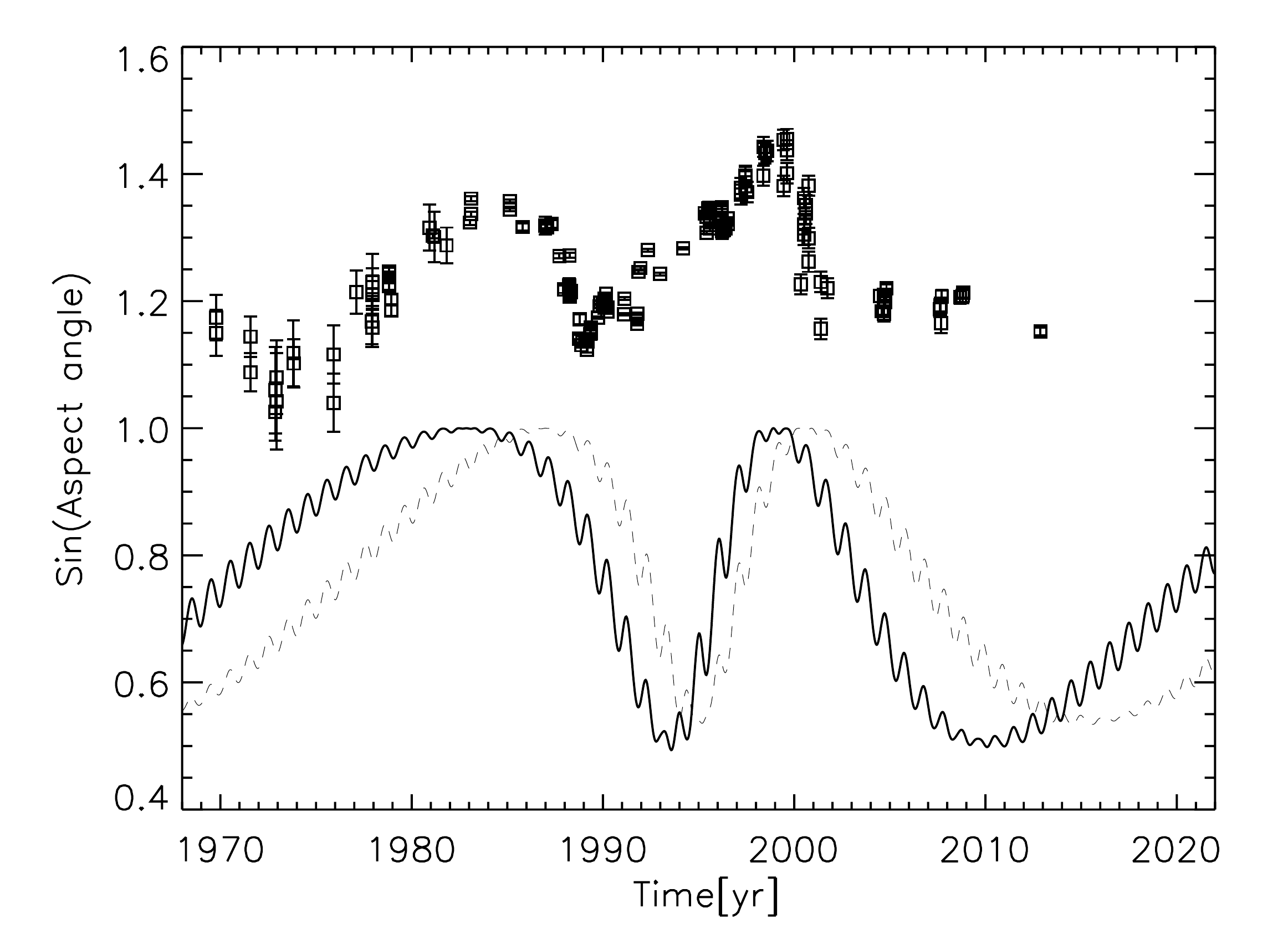}
\caption {\textit{Coincidence of the brightness minima with aspect angle near 90 degrees}: In continuous line we show the sine of the aspect angle of Chiron as a function of time using the preferred pole solution ($\lambda$=$144\degree$, $\beta$=24$\degree$). The dotted line corresponds to the other solution for the pole direction ($\lambda$=$352\degree$, $\beta$=37$\degree$). The square symbols represent the absolute magnitude of Chiron from \cite{Belskaya2010}  divided by a factor 5 for an easier viewing. As can be seen there is coincidence between aspect angle and absolute magnitude, but there is a shift between the aspect angle maxima and the absolute magnitude maxima for the $\lambda$=$352\degree$, $\beta$=37$\degree$ pole solution. }
\label{fig5}
\end{figure}

Using a compilation of absolute magnitude measurements of Chiron at different epochs from \cite{Belskaya2010}, we realized that the two deep maxima in absolute magnitude (brightness minima) coincide with the epochs when the  proposed rings were edge-on with respect to an Earth observer (aspect angle of 90 degrees). In  Fig. \ref{fig5} we plot the absolute magnitude and the sine of the aspect angle of the rings for the two pole solutions, as a function of time. We do this to illustrate when the rings are edge-on (maxima of the curves) and to compare the maxima in the curves with the absolute magnitude measurements (squares). As can be seen, the second pole solution ($\lambda$=(144$\pm$10)$\degree$, $\beta$=(24$\pm$10)$\degree$) gives a better match to the times when the maxima in absolute magnitude (brightness minima) are reached. Our explanation of this coincidence is that the rings have an important effect on the long term brightness of Chiron's system, and we built a simple model of the brightness of Chiron's main body and its ring system, in which a large fraction of the brightness comes from the rings. A similar model was built to successfully explain Chariklo's absolute magnitude changes over time \cite{Duffard2014b}.

With the second pole direction ($\lambda$=(144$\pm$10)$\degree$, $\beta$=(24$\pm$10)$\degree$), an overall match to the absolute magnitude measurements of Chiron (except the parts with large brightness excursions due to activity outbursts) can be obtained.  Activity outbursts were observed in 1989 \citep[e.g.][]{Meech1989} and in 2001 \citep{Romon-Martin2003}. Also, \cite{Bus2001} suggest that an outburst could have taken place around the 1970's. The key feature of the model is the inclusion of the changing aspect of Chiron's rings.

The model is very simple and takes into account the total flux coming from both Chiron's nucleus and its rings:

%\begin{equation}
%$$ \frac{F_{tot}}{ F_{Sun}} = A_p  p_V f(\alpha)+ \frac{\mu_0}{\pi}+\frac {\omega_0} {4(\mu + \mu_0)} P(\mu, \phi, -\mu_0, \phi_0)
%(1-exp^{-(\frac{1}{\mu} +\frac {1}{\mu_0}) \tau }) 2 \pi W a \mu +
%\frac{\mu_0}{\pi}+\frac {\omega_0} {4(\mu + \mu_0)} P(\mu, \phi, -\mu_0, \phi_0)
%(1-exp^{-(\frac{1}{\mu} +\frac {1}{\mu_0}) \tau_2 }) 2 \pi W_2 a_2 \mu
%$$
%\end{equation}

%ecuacion escrita por Rene
%\begin{equation}
%$$ \frac{F_{tot}}{\pi F_{Sun}} = A_p  p_V f(\alpha)+ \frac {\omega_0
%(1-exp^{-(\frac{1}{\mu} +\frac {1}{\mu_0}) 2 \tau }) 2 \pi W a \mu}  {8
%}+ \frac {\omega_0 (1-exp^{-(\frac{1}{\mu} +\frac{1}{\mu_0}) 2 \tau_2 }) 2
%\pi W_2 a_2 \mu}  {8 } \eqno{1}$$
%\end{equation}

%\begin{equation}
%$$ \frac{F_{tot}}{ F_{Sun}} = A_p  p_V f(\alpha)+ \frac {\omega_0} {4 \pi} 2 \pi W a \mu +
% \frac {\omega_0} {4 \pi} 2 \pi W_2 a_2 \mu
%$$
%\end{equation}

\begin{equation}
$$ \frac{F_{tot}}{ F_{Sun}} = A_p  p_V f(\alpha)+ p^{Ring1}_V f'(\alpha) 2 \pi W_1 a_1 \mu +
 p^{Ring2}_V f'(\alpha) 2 \pi W_2 a_2 \mu
$$
\end{equation}where $F_{tot}$ is the flux received at Earth, $ F_{Sun}$ is the solar flux at Chiron, $A_P$ is the projected area of Chiron's nucleus,
$p_V$ is the geometric albedo of the main body, $f(\alpha)$ is the solar phase function, $p^{Ring1}_V$ is the albedo of the first ring,
$f'(\alpha)$ is the solar phase function of the ring,
%$P(\mu, \phi, -\mu_0, \phi_0)$ is the single scattering phase function where $\phi$ is the azimuth angle of the observer, $\phi_0$ is the azimuth angle of the sun, $\mu_0$ is the cosine of the solar zenith angle from the rings,
$\mu$ is the absolute value of the cosine of the observer zenith angle
from the rings  (which coincides with the cosine of the rings' aspect angle) and
%and $\tau_i$ is the optical depth.
$2 \pi W_i a_i \mu$ is the projected area of the rings, where $W_i$ is the radial width of
the $i$ ring, and $a_i$ is the radial distance of each ring to the main body.
%The observer zenith angle from the rings is almost identical to
%the aspect angle and here we can take the approximation that $\mu$=$\mu_0$ because the sun and the observer are almost in the same direction
%as viewed from the distant Chiron.

%However, that simplistic model based on the single scattering approximation for the rings is not valid because the optical depth of the rings is at least as high as that for Chariklo. With an optical depth of $\sim$ 0.4 \citep{BragaRibas2014}, the single scattering approximation is not justified, specially at very small values of $\mu$ for which the line of sight optical depth would diverge. This is even more true because we think that Chiron's rings are even optically thicker than those of Chariklo. Hence, multiple scattering computations would be needed, but because we do not know most of the intervening parameters, we decided to simplify the approach by just assuming isotropic reflection from the rings. Hence we assume that the solar flux falling on the rings is spread over 4$\pi$ steradians. With this simplified model we can still gather some useful insight on the rings properties needed to explain the overall photometric behavior of Chiron. In this case, the flux from Chiron and its ring system is:

For $f(\alpha)$, the solar phase function, we take it equal to 1 for all phase angles, which is a valid approximation
because we know that the phase angle dependence is weak  (around a few cents of magnitude per degree according to \citet{Belskaya2010}) and the phase angle ($\alpha$) in Centaurs is only a few degrees at most. We assume the same for the rings phase function.
%The parameters $\tau_i$, $W_i$ were initially assumed to be similar to those of the Chariklo rings
%from \cite{BragaRibas2014}.
%We also take the albedo of the rings as 0.08, which is the albedo determined for the rings of Chariklo from spectroscopic modeling in \cite{Duffard2014b} and $a_i$ was taken as 324km from the occultation results.

We take the albedo of the rings as 0.17 from the spectroscopy constraints, and $a_i$ was taken as 324 km from the occultation results.
The geometric albedo of Chiron's main body, $p_V$, was taken as 0.08 from \cite{Lebofsky1984} because the observations in that work were performed
when the ring was almost edge on so the albedo determination
was not contaminated by the rings. Besides, the surfaces of the centaurs are thought to be like those of the comets, whose geometric albedo is typically around a few per cent. Comets and Centaurs are dynamically linked so it seems reasonable that both populations have similar albedos. 
%from \cite{Lebofsky1984} because the observations in that work were performed when the ring was almost edge-on so the albedo determination was not contaminated by the rings.
It is important to note that $\mu$ for the different epochs depends on the rings pole direction.  Also, the projected area $A_p$ was computed by assuming that Chiron is a triaxial ellipsoid with the following dimensions: $a$=100 km, $b$=84 km, $c$=70 km, where $a,b,c$ are the semimajor axes of the triaxial ellipsoid describing the body. These dimensions are obtained by using the $a/b$ axial ratio determined for Chiron by \cite{Groussin2004} and putting the constraint that the equivalent diameter of a sphere is $\sim$ 180 km. We use 180 km because this is an average value among all the values reported in the literature as already mentioned in previous sections. The axial ratios were also used in the subsequent section to explain the amplitude of the rotational lightcurves. The model does not take into account the area of the rings occulted by the main body, but this is a tiny correction and would only affect a narrow range of aspect angles.

\begin{figure}
\includegraphics[width=9cm, angle=0]{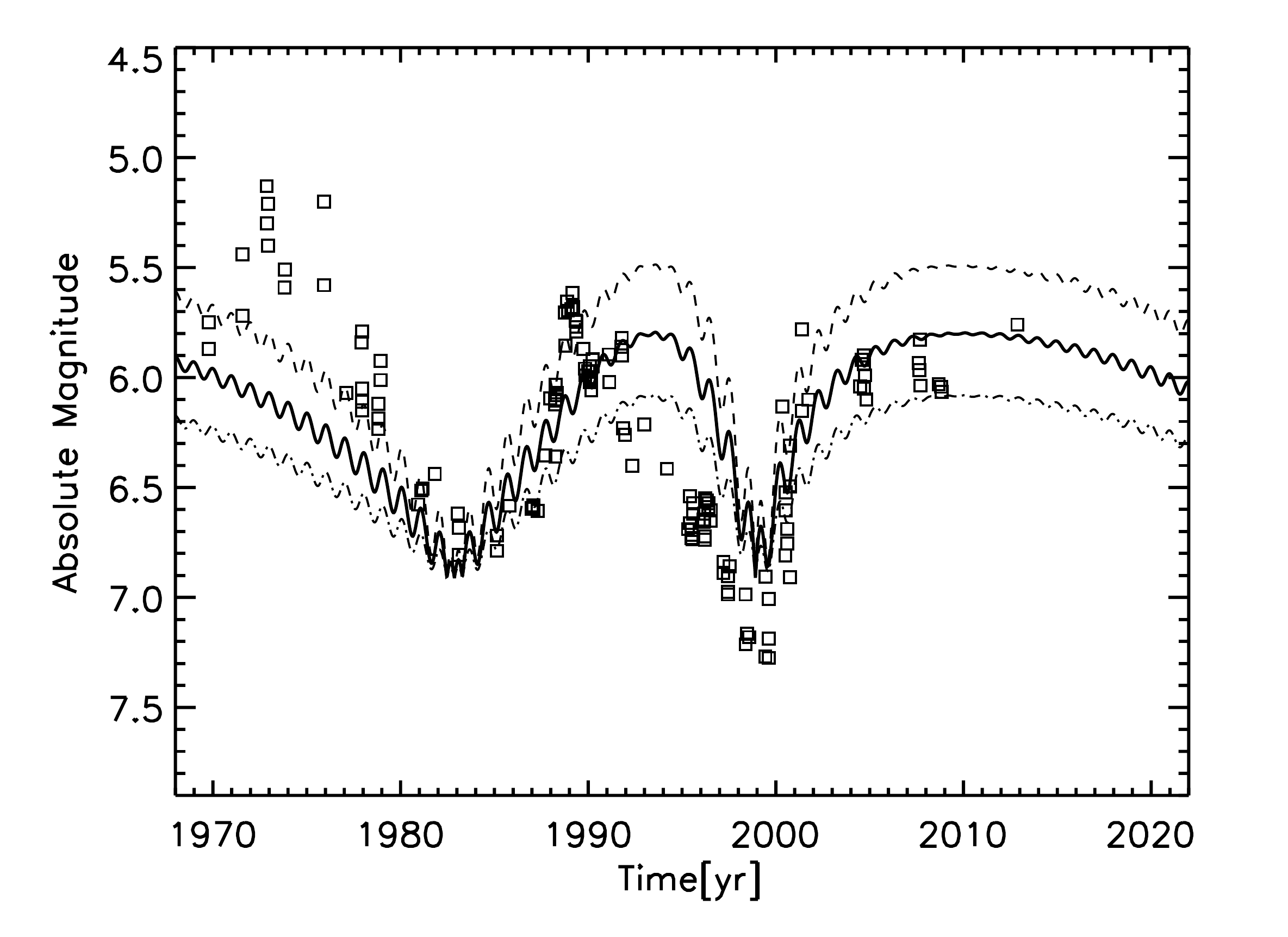}
\caption {\textit{Model to explain the absolute magnitude in V band}: In continuous line we show the results from the nominal model that can explain the overall photometry of Chiron from \cite{Belskaya2010} (square symbols). The upper and lower dashed lines represent models with ring albedos of 0.1 and 0.27 respectively.}
\label{fig6}
\end{figure}

In Figure \ref{fig6} we show the brightness of Chiron and its rings from the model together with the measurements of absolute magnitude of Chiron at different epochs compiled by \citet{Belskaya2010}.
 These data are mostly comprised of the data by \cite{Duffard2002} and \cite{Bus2001}

 In this model there are no free parameters because all the parameters are obtained from either the literature or from our results of the previous sections. However, given the uncertainty range in some of these parameters, it is useful to test the effect of some of them. The parameter that has the largest relative uncertainty is the geometric albedo of the rings, so we have tested its effect by using a 0.1 albedo and a 0.27 albedo in addition to the nominal 0.17 value derived from the spectroscopy.
%The parameters that have the largest effect in altering the shape of the curve are the albedo of the rings together with the width of the rings and the orientation of the rings pole. The size of Chiron and its albedo play a small role too and the axial ratio of the body has almost no effect.

\begin{figure}
\includegraphics[width=9cm, angle=0]{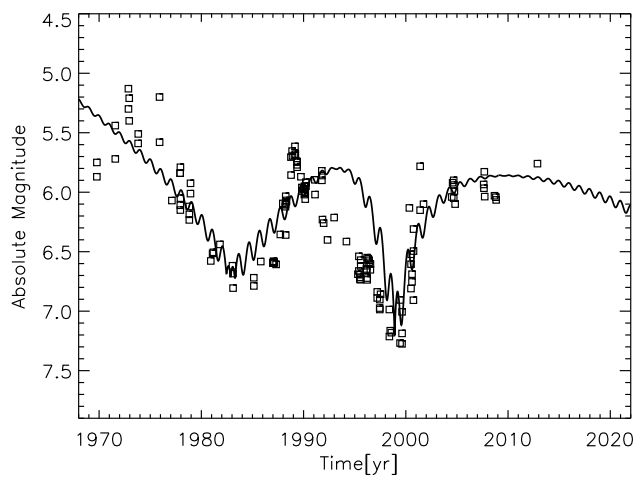}
\caption {\textit{Model that includes a vanishing coma to explain the absolute  magnitude in V band}: In continuous line we show the results from the model discussed in the text but including the contribution of an exponentially decaying coma. The observations as compiled in \cite{Belskaya2010} are shown as square symbols.}
\label{fig7}
\end{figure}

 The lower dashed line in Figure \ref{fig6} corresponds to a model with a geometric albedo of 0.1 for the rings. In thick line we show the model with the nominal parameters (geometric-albedo of the rings 0.17), which provides a better agreement with the observations. In the upper dotted-dashed line we present a model with a geometric albedo of 0.27 for the rings. It is important to note that the contribution from the rings to the total visible flux is similar to that of the projected area of Chiron (except when the rings are seen nearly edge-on, at aspect angles around 90 degrees, when the rings do not contribute). This can explain a greater than one magnitude difference from maxima to minima. Chiron's exact shape is not a relevant factor to explain the overall absolute magnitude measurements. A ring system is the only way that we can think to cause a large magnitude change of around 1 or more magnitudes. As already stated, the model fails to reproduce the brightness local maxima that are related to activity outbursts. We know that Chiron has been an active body and outburst episodes are unpredictable. Besides that limitation, the model does not completely reproduce the absolute magnitude of the observed brightness minima because the observations show that the two brightness minima have different values. We can explain a deeper brightness minimum in 2000 than in 1983 if the albedo of Chiron had decreased significantly in the $\sim$20 years from 1983 to 2000, which would be difficult to explain (unless there is some dark dust recoating the nucleus or extinguishing its signal). The same would be true if the size of Chiron had changed considerably in $\sim$20 years, which is even more difficult to justify from a physical point of view. But there is an alternative way to explain the difference in the two minima and we can easily model it. If we add the contribution of a cloud of debris or dust whose brightness decays exponentially with time, with a time constant of around 10 years (see Fig. \ref{fig7}), we can explain the different depths of the two observed brightness minima.  The only other parameters in the model are the albedo of the coma particles, which we set to 0.06 (a typical value of comet dust albedo), and the initial cross section of the cloud of particles, which we determined iteratively to be $\sim$ 85000 km$^2$. This area is comparable to the total cross section of Chiron and its rings.  Mathematically, the modified model can be expressed with the following equation:

\begin{equation}
\begin{split}
\frac{F_{tot}}{ F_{Sun}} = A_p  p_V f(\alpha)+ p^{Ring1}_V f'(\alpha) 2 \pi W_1 a_1 \mu + \\  
   p^{Ring2}_V f'(\alpha) 2 \pi W_2 a_2 \mu + A_c p_d\exp{(-(t-t_0)/\tau_d) } 
\end{split}
\end{equation}where $A_c$ is the initial scattering cross section of the cloud, $p_d$ is the geometric albedo of the dust, $\tau_d$ is the decay time, $t$ is time and $t_0$ is the initial time at which the dust cloud starts to decay. 

Because there are hints for plenty of debris or dust around Chiron, it seems plausible that some debris may be escaping from Chiron with an exponential behavior and therefore the cloud brightness could decay exponentially.  \citet{Elliot1995} proposed that Chiron could be surrounded by a bound coma to explain some of their observations. It seems possible that such a bound coma or cloud is not completely bound in time scales of tens of years and some material is escaping slowly. An exponentially decaying coma has already been proposed for Chiron by \cite{Meech1990} and \cite{Duffard2002}, although with different time constants.  Also, it is important to note that the absolute maximum in the photometry around 1973 is best explained in this scenario. In this modified model we set the start of the decay $(t_0)$ at the first year for which we have data (1970), but the decay might have started in 1973. This would provide an even better agreement with the photometry, but at the expense of adding more free parameters to the model, like the time of outburst and the cloud growth parameters, for which we do not really have good constraints. Many sophistications to the model are possible, including resurfacing or coating mechanisms, and modeling of the outbursts, but they are not straightforward to implement. Here we just presented some rough ideas. In summary, the model whose key feature is a ring system with changing cross section is consistent with the overall behavior of Chiron's brightness.

\section{Lightcurve amplitude}

\begin{figure}
\includegraphics[width=9cm, angle=0]{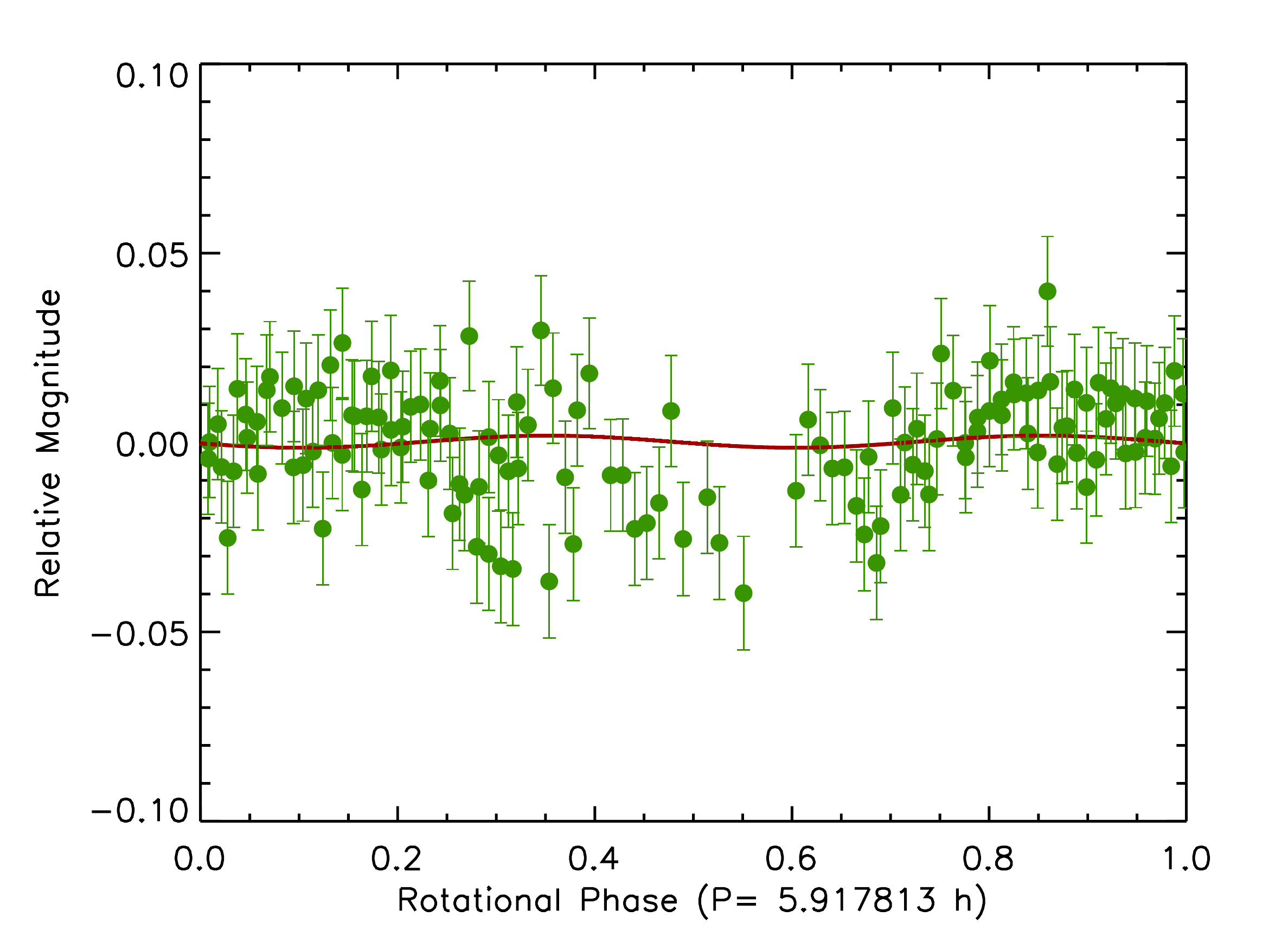}
\caption {\textit{Lightcurve of Chiron}: Rotational lightcurve obtained from the CAHA 1.2m telescope in 2013. A two term fit (two oscillations per rotation cycle) is shown on top of the data. The Julian Date for zero phase angle is 2451783.41511.}
\label{fig8}
\end{figure}

We got 250~s R-band exposures of Chiron with the 4k$\times$4k DLR-MKIII CCD camera of the 1.23~m Calar Alto Observatory telescope in Almeria, Spain.  The field of view of the instrument was 21.5 x 21.5 arcmin and the plate scale was 0.63 arcsec per pixel. The images were obtained on October 9th and 10th, 2013. The telescope was tracked at the sidereal rate. Bias frames and twilight flatfield frames were taken each night to calibrate the exposures. No sign of coma was observed in the images. The profiles of Chiron were entirely identical to stellar profiles, so no signs of activity were detected. From the images we obtained time series relative photometry using the same techniques as described in \citet{Thirouin2012}. We used several synthetic aperture radii and chose the results that gave the lowest scatter. The final time series relative photometry is given in online Table~\ref{tab:photo}. From the photometry we derived the rotational lightcurve by phasing the photometry to the known 5.917813~h period of Chiron \citep{Bus1989,Marcialis1993}. We fitted a 2-term series to the lightcurve, because we know that the lightcurve is double-peaked \citep{Bus1989,Marcialis1993}. The derived lightcurve amplitude turned out to be 0.003~mag $\pm$ 0.015, much smaller than the 0.088~mag amplitude derived by \citet{Bus1989} in 1986 and 1988. The lightcurve and the fit are shown in Figure~\ref{fig8}.
%ojo, acabo de ver que Fornasier et al. 2013 incluyen fotometria de Chiron obtenida en el 1.2m de CAHA en diciembre de 2011, imagino que por Mottola, y dicen que la amplitud de la curva de luz es de 0.06 a 0.07 magnitudes, pero muestran una curva de luz faseada (folded) a un periodo de 5.4h, que no es el correcto, tienen muy pocos puntos, sin barras de error, y no estiman tampoco el error en la amplitud que ellos mencionan... Menos mal que no soy coautor de ese art�culo... Los coautores, que se pronuncien...

\begin{figure}
\includegraphics[width=9cm, angle=0]{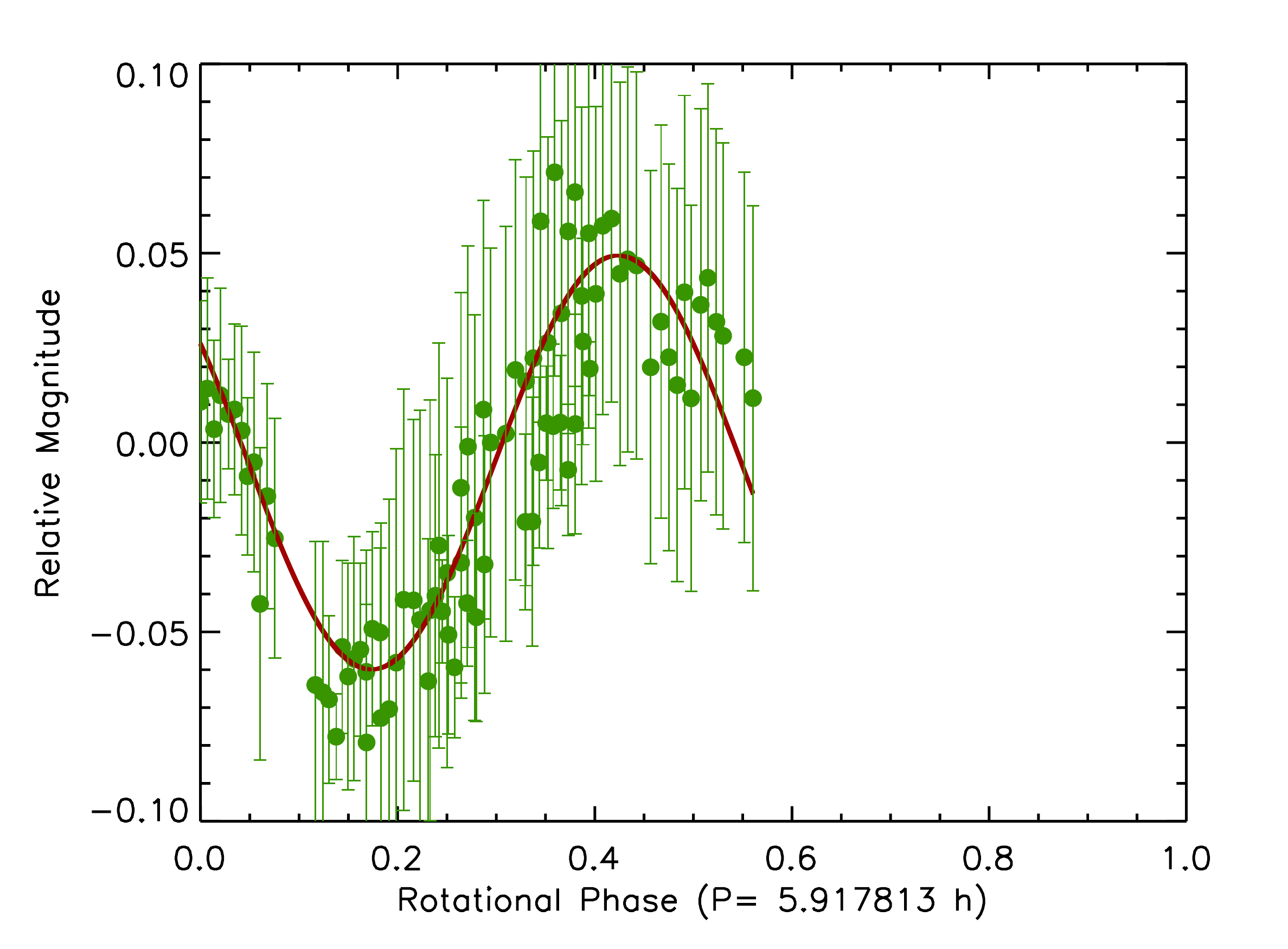}
\caption {\textit{Lightcurve of Chiron}: Rotational lightcurve observed from CASLEO 2.15m telescope in 2000. A two term fit (two oscillations per rotation cycle) is shown on top of the data. The Julian Date for zero phase angle is 2451783.41511.
%The lower panel shows the residuals of the fit.
%\textbf{falta la e a magnitud. Porque hasta 0.5 en fase rotacional? No se si esta es la curva de caha o de casleo? }
}
\label{fig9}
\end{figure}

We also derived time series photometry of Chiron from CCD images that were obtained at the CASLEO 2.15~m telescope on August 26th, and 28th, 2000, but had not been analyzed. The images were obtained with a TK 1K CCD camera in the R-filter and consisted of 60~s to 100~s exposures. The field of view of the instrument was 4.6 x 4.6 arcmin and the plate scale was 0.54 arcsec per pixel. The telescope was tracked at sidereal rate. Bias frames were available, but flatfield frames were not, so we applied no flatfield correction. This resulted in slightly poorer photometry than what would have been possible, but the results are still useful. We derived the time series photometry by using the same techniques as for the 1.23~m telescope images. The results are given in Table~\ref{tab:photo} and the rotational lightcurve is presented in Figure~\ref{fig9}. In this case the observations did not cover a full rotation of Chiron, but were enough to cover a substantial part of it and allowed us to fit an amplitude of 0.097 $\pm$ 0.039~mag.

We have modeled the shape of Chiron's nucleus as a triaxial ellipsoid, which is a first order approximation. The cross section variations cause the main part of the rotational variability. For a triaxial ellipsoid, the amplitude of the lightcurve ($\Delta m$) is, according to \citet{Binzel1989}:

\begin{equation}
$$
\Delta{m} = 2.5~log\left( \frac{a}{b}\right)  - 1.25~log\left(
\frac{a^{2}\cos^{2}\xi + c^{2}\sin^{2}\xi}{b^{2}\cos^{2}\xi +
c^{2}\sin^{2}\xi}\right)
$$
\end{equation}where $\xi$ is the aspect angle and \textit{a}, \textit{b}, and \textit{c} are the semi-axes lengths of the ellipsoid with $a>b>c$.  This kind of model has already been used to determine the pole of Pholus, another centaur \cite{Tegler2005}. The aspect angle is related to the ecliptic longitude and latitude of the spin vector ($\lambda_p , \beta_p$) by standard equations \citep[e.g.][]{Tegler2005}.

However, this simple model, which is commonly used to model asteroid's lightcurve amplitudes, does not take into account the brightness of rings. As pointed out before, this contribution is significant, and attenuates the amplitude of the lightcurve whenever the rings are not edge-on. We have computed the decrease of the amplitude as a function of time by adding the rings flux contribution as in equation 1.

\begin{figure}
\includegraphics[width=9cm, angle=0]{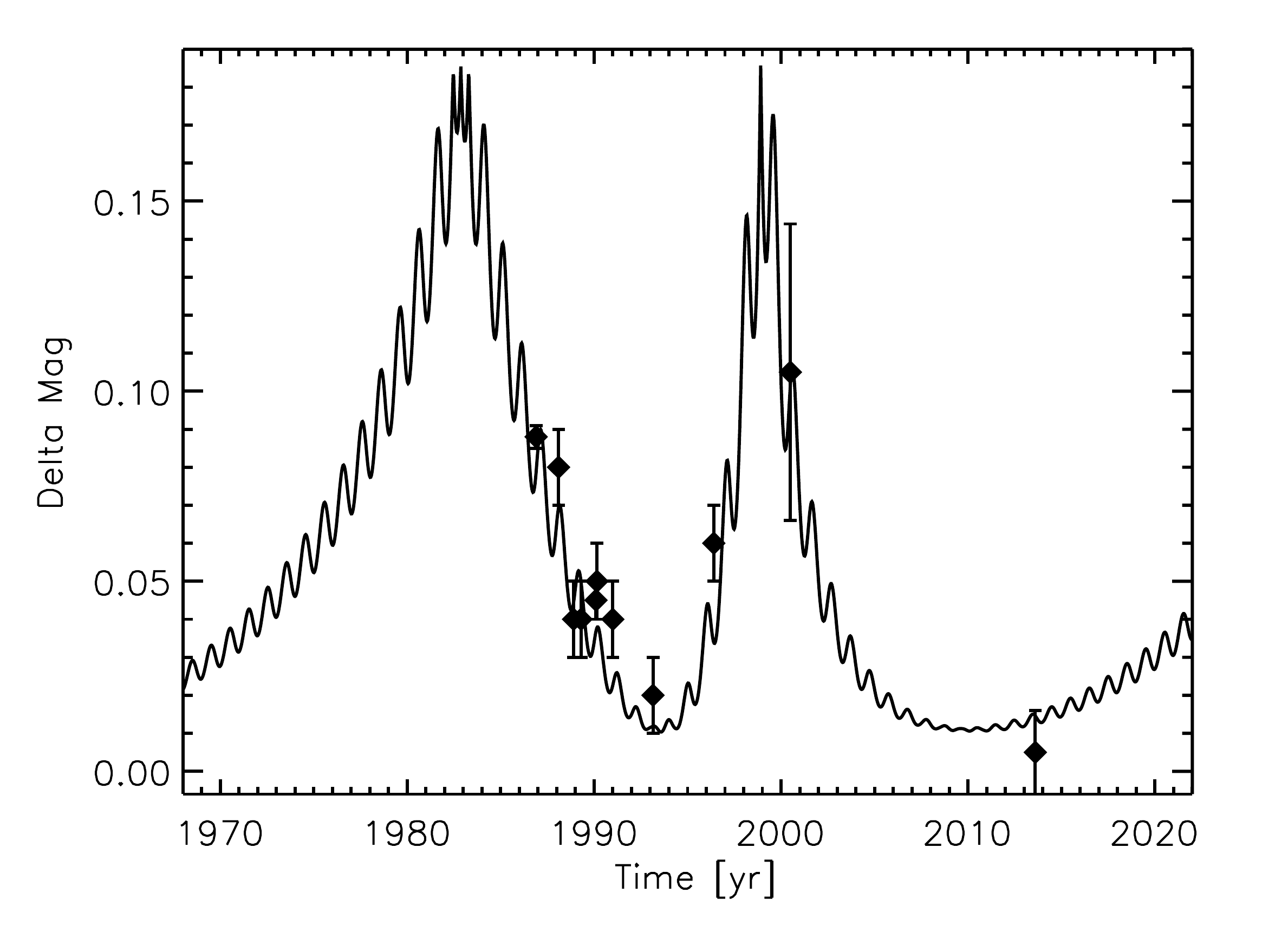}
\caption {\textit{Amplitude versus time}: In continuous line we show the amplitude of the rotational lightcurve for an assumed triaxial shape of Chiron (with axial ratios a/c=1.43 and b/c=1.2) and using the nominal pole direction determined from the stellar occultations ($\lambda$ $\sim$ 144$\degree$, $\beta$ $\sim$ 24$\degree$). The diamond symbols represent the observed amplitudes for the specific dates of observation (see text). As can be seen, the model provides agreement with the observations.}
\label{fig10}
\end{figure}

Using axial ratios \textit{a/c} and \textit{b/c} of 1.43 and 1.2 respectively and the preferred pole direction $\lambda_p$=144$\degree$, $\beta_p$=24$\degree$ derived from the occultation, we can explain the lightcurve amplitude of the \citet{Bus1989} data (which corresponds to data obtained in 1986 and 1988), the amplitude in subsequent years compiled by \cite{Groussin2004} and our new results obtained in the present work. All this is depicted in Figure~\ref{fig10}. Note that the value of the \textit{a/c} axial ratio is rather irrelevant, because the lightcurve amplitude basically depends on the \textit{a/b} ratio alone. We chose a ratio \textit{a/c} rather arbitrarily, although it is consistent with the occultation of the main body. Other values of \textit{a/c} can also result in good fits provided that the axial ratio \textit{b/c} is chosen so that \textit{a/b} is around 1.16.  This $a/b$ axial ratio of 1.16 $\pm$ 0.03 was found by Groussin et al. (2004). It is also important to note that some degree of rotational variability can be due to albedo variegations on Chiron's surface. Hence, the model is a first order approximation.  Nevertheless, the agreement of the model with the observations is remarkable. The reduced $\chi^2$ of the fit is 1.9. The pole solution  $\lambda_p$=352$\degree$, $\beta_p$=37$\degree$ does not produce a satisfactory fit to the observed amplitudes.

\section{Other arguments or indications supporting the existence of a ring system}

According to \citet{Ruprecht2013a} the very sharp occultation features in 2011  were symmetric jet-like features that could be caused by a near-circular arc or shell of material ejected by Chiron and expanding away from the nuclear surface produced by one or more active regions on the nuclear surface. But a shell would cause an occultation along the whole chord from the start of the shell to the end of the shell, not sharp isolated features. One can argue that the sharp features could be caused by the increased apparent optical depth of a narrow spherical shell at nearly tangent angles from the observer, but at least some degree of occultation should be detectable along the chords and this was not the case. Besides, the shell would have to be ellipsoidal, not spherical, to explain the features detected from the two observing sites, when they are projected in the sky plane.  To our knowledge, no ellipsoidal shells have ever been observed in a comet. And to make matters worse, there should be two concentric narrow shells with a small gap in between. Such narrow shells have never been observed in cometary coma. Hence, the shell explanation has many problems. Regarding jets, we cannot explain why they would give symmetrical configurations and again, they should give rise to a double structure with a small gap in between.  Also, it must be pointed out that Chiron was in a quiet state at the epochs of the occultations; it was not active. It suffered no outburst and did not show coma in 1993, 1994 and 2011. Hence, apparently there was no source from which a jet or a shell could emanate. In addition to all this, a jet or a shell cannot explain the sharp minima in Chiron's long term brightness. Therefore, not only are the rings a good explanation to all the features mentioned in the previous section, but alternative scenarios such as a shell or a jet fail to explain the observed phenomena.

The presence of rings can be related to the considerable scatter in the values of the equivalent diameter and albedo of Chiron reported in the literature from thermal observations, because the rings contribution could be important, at least in the optical, and has a strong dependence with the aspect angle.
 It is also remarkable that the geometric albedo of the rings determined from the spectroscopic constraints is consistent with the geometric albedo of the rings that is used in the photometry model to explain the long term brightness behavior of Chiron.
 On the other hand, the widths of the extinction features, their separations and their optical depths are, as already mentioned in the introduction, nearly identical for Chiron and for Chariklo. These are clear indications that Chiron extinction features are probably caused by the same type of structure as in Chariklo.
Also, the radius of the rings can be a further indication: The radius of the  proposed ring system in Chiron from our results is 324$\pm$10~km, around a factor 0.82 smaller than that of Chariklo,  which is around 400 km \citep{BragaRibas2014}. This seems reasonable given that the size of Chiron is smaller than that of Chariklo. For plausible densities of both Chiron and a satellite, the ring system is within the Roche limit in both Chariklo and Chiron. Therefore, both Chariklo and Chiron would have ring systems with similar features. Nevertheless, as already stated, Chiron's ring system  would be brighter compared to Chiron than Chariklo's rings compared to Chariklo. Besides, Chiron may have dust or debris in orbit, not just a ring system. All this might indicate that  the proposed ring system  could be younger than that of Chariklo, if the rings are formed from collisional or satellite disruption processes. See discussion section.

%The overall shape of our model is a slightly deformed maclaurin spheroid. Given that the rotation period of Chiron is known we can compute the density of a hydrostatic equilibrium maclaurin body that would have an axial ratio around 1.4 and a rotation period of $\sim$5.91h. The resulting density is 1400 kg/m$^3$. We think that Chiron is probably too small to be in hydrostatic equilibrium, but if it is a rubble pile, its shape could be consistent to first order with that of a hydrostatic equilibrium body. In that case, the derived density would be a first order guess.

\section{Discussion}

The fact that not only Chariklo but Chiron, the second largest centaur after Chariklo,  can also have a ring system is remarkable. This might be a hint that the origin of the rings could be related to phenomenology specific of centaurs, or perhaps rings are common in the outer solar system from Jupiter and beyond. Also, as is the case for Chariklo's ring system, Chiron's ring system has a gap and the rings seem to have sharp edges. Hence, both ring systems might be confined by shepherd satellites, as proposed in \citet{BragaRibas2014}.

The rings in Centaurs may come from material released from the main body after an impact, but the rings may result also from the destruction of a small satellite or from the release of material from a satellite upon impact.  If the rings origins have to do with collisions on the main body smaller centaurs (which have a smaller cross section) would have a smaller probability of developing rings. This might be tested in the future if we can observe occultations by Centaurs of different sizes. In these latter two cases the rings would result from two-step processes: first an event that generates one or several satellites and then a collision on a satellite: Whether the satellite or satellites result from ejecta of direct collisions to the main bodies or result from other phenomena (for instance the satellites were captured while the body dwelled in the trans-Neptunian belt or the satellites were formed from rotational fissions) needs to be investigated as well.

Several studies indicate that satellites may be very common in the trans-Neptunian belt and more common the closer to the primary object \citep{Noll2008}. They can come from very different formation scenarios. Even though capture is in general the favored mechanism, for small satellites close to the primaries the main causes can be collisions or even primordial rotational fissions of the primaries. A shattering collision with a satellite in the lifetime of a centaur releasing material equivalent to a 1~km-radius body would occur only if the current flux of impactors in the centaur region is at least one or two orders of magnitude higher than what is derived in \citet{Levison2000}. Such an enhanced flux rate with respect to \citet{Levison2000} would be consistent with the fact that the impact of a nearly km-sized object has been observed on Jupiter after the Shoemaker-Levy 9 impact and this requires a larger impact flux than previously thought \citep{Sanchezlavega2010}. Hence, under this view, it appears that the impact rate in the outer solar system would have been underestimated in the past.

The rings might also come from sublimation-driven activity events on the satellites, but this would require that the satellite or satellites had more volatile material than the primary itself. Besides, fine grained material, which is typical of cometary activity would easily disperse in small time scales due to solar radiation and would not easily form a long-lasting ring. Therefore, sublimation-driven activity on a satellite does not seem to be a very plausible cause for the rings material.

Thus far several occultations of bright stars caused by large TNOs  have been recorded: from historical occultations by Pluto \citep{Pasachoff2005,Elliot2003,Elliot2007,Person2008,Young2008} to occultations by Varuna and a few others \citep{Sicardy2010,Sicardy2011,Ortiz2012,BragaRibas2013}. For some of these occultations the stars involved were bright enough so that very good signal to noise was achieved. Hence, sharp secondary events caused by rings should have been recorded already if rings were very common in trans-Neptunian Objects (TNOs). This could mean that the rings in TNOs are not common and under this interpretation rings  of Centaurs would not be formed in the trans-Neptunian belt. But the lack of detection of rings in TNOs could also be due to size issues, because we have recorded occultations of only the largest TNOs. Except for the occultation of (55636) 2002 TX$_{300}$\citep{Elliot2010}, which is a very peculiar TNO belonging to the so-called "Haumea family", and is not very representative of most of the TNOs \citep{Licandro2006}, there are no recorded occultation events of TNOs in the size range of Chariklo and Chiron  (180 km to 260 km in equivalent diameter). The smallest and representative TNO whose occultation has been recorded is 2002 KX$_{14}$ \citep{Alvarez-Candal2014}. In that work an equivalent diameter between 365 and 455 km is reported but no secondary events were observed despite the  0.265 s time resolution, which would have allowed the detection of rings of similar optical depth to those reported for Chariklo. And even if  (55636) 2002 TX$_{300}$  is not a very representative TNO, it must be mentioned that no brief secondary events were detected by \citet{Elliot2010} despite the very high time resolution ( 0.072 s) and good signal to noise ratio  (105 for a hypothetical 1s integration). Their Figure 1 clearly shows that the best lightcurve does not have brief sharp events prior or after the main occultation.

Therefore, there is a slight indication that rings may be formed only in the Jupiter to Neptune solar system region where centaurs reside rather than in the trans-Neptunian region. This could perhaps be due to particularly enhanced collisional rates compared to the trans-Neptunian region. Under this view the rings would form from collisions on the main body or on a satellite that was either primordial or captured while the object was in such a region.

One can wonder why rings have not been identified in main belt asteroids despite there have been many occultations observed with good time resolution. The lack of rings in the main asteroid belt may be related to the average impact speed in the asteroid belt compared to the impact speeds in the regions where the centaurs reside. But this is too speculative at the current time. Outside the main asteroid belt, in the Jupiter trojan's reservoir of small bodies, there are fewer occultations observed in comparison with the main asteroid belt.  It may be interesting to point out that Agamemnon ((911) or 1919~FD), a large Jupiter Trojan, showed a secondary and sharp event during an occultation \citep{Timerson2013}. This event has been interpreted as an indication of a small satellite of a few km, but perhaps it is revealing the presence of a ring that was not detected in the other occultation chords simply because of insufficient signal to noise. The secondary event was recorded only by the observer who used the largest instrument, a 0.36~m telescope, whereas the other observers used 5~cm binocular video systems, too small to detect secondary events of this sort. But a ring should have produced two symmetrical secondary events, not just one.

%The observer with the 0.36~m telescope should have found a symmetrical previous sharp event several seconds before the main drop if the secondary event were produced by a ring, but this was not the case; the observer detected only a secondary event, not two.

Therefore, so far the best explanation seems to be a satellite, although a partial or incomplete ring (an arc) should not be discarded yet. It must be noted that Agamemnon has large excursions of its magnitude as reported to the Minor Planet Center compared to the predictions from JPL ephemeris, much like Chiron's case. This large variability could be caused by a partial ring. We have done a search for Trojan asteroids showing brightness excursions in a similar way as \cite{Cikota2014} did to find main belt comet candidates. The only trojans clearly showing large variations were Agamemnon, Achilles and Aneas.

Concerning the historical absolute magnitude of Chiron versus time, we can explain the brightness minima and the overall behavior with the ring system, which is remarkable. So far there was no model that could explain the changing absolute magnitude of Chiron. Of course, there are details that cannot be fit by our model, particularly at the epochs of Chiron's activity outbursts, when episodes of sudden dust release take place. Concerning the activity outbursts one could think that there are sublimation-driven active regions within the rings or at a high latitudes on Chiron's surface that get activated when the solar incidence angle is appropriate. However, as we show in the aspect angle plots versus time, the same aspect angles as those in the outburst epochs (for example 1988-1989 and 2001) repeated at other epochs but there were no activity outbursts in those years. Hence the activity cannot be completely controlled by insolation factors. From our study we can propose other scenarios that would give rise to the transient development of coma. A possible scenario could be the fall back of debris that has not been accreted to the rings, or the fall of material from the rings. The impact of debris on the surface of Chiron might suddenly release dust at unpredictable times. These outbursts would not be sublimation-driven, contrary to the usual comet outbursts.  If the debris follows a size distribution, it may be possible that small outbursts caused by debris of small size could be more frequent than the two outbursts detected thus far and these smaller outbursts would be detectable in accurate time series photometry measurements. In summary, we are proposing that activity outbursts could be caused by the fall back of material that may be part of the ejecta of an old collision of a small body with Chiron or with a satellite. Part of the ejecta could form a disk that could evolve to a ring and part of the ejecta could eventually fall back and cause transient comae.

Finally, we would also like to speculate on another interesting topic. An intriguing possibility is whether the color bi-modality observed in the color distribution of centaurs \citep{Peixinho2003, Peixinho2012} is due to centaurs with rings and centaurs without rings. Centaurs with rings would have a bluer or a more neutral color than the other centaurs because of the spectral contribution of the rings, which is different to that of the main body. Indeed, the centaur color distribution shows a group with red colors and a group of more neutral colors (among which Chiron and Chariklo are). If the bi-modality is caused by the presence of rings, rings would be very common in centaurs. Recently, \citet{Peixinho2012} have proposed that the color bi-modality is also present in small TNOs. Hence, if the color bi-modality and the presence of rings are related, small TNOs might have rings as well and under this view the origin of the rings would most likely be in the trans-Neptunian region rather than in the Jupiter-Neptune region where the centaurs currently reside.

\section{Conclusions}

We presented occultation, photometric, and spectroscopic observations which indicate the presence of ring material around Chiron. In particular, the 2011 occultation lightcurve obtained by \citet{Ruprecht2013a} shows secondary events having similar characteristics to that recently found for Chariklo, another centaur. From the occultation an approximate diameter for the  proposed ring system of 324 $\pm$ 10 km has been derived and two possible pole orientations have been determined. We favor the pole solution $\lambda$ $\sim$ 144$\degree$, $\beta$ $\sim$ 24$\degree$ because it can better explain the rotational lightcurve amplitude at several epochs and can also explain the overall historic photometric behavior of Chiron (its absolute magnitude versus time). Also, the changing orientation of the  proposed ring system is consistent with the observed variability in the water ice spectral features.  Even though the occultation evidence alone might not seem conclusive enough, 
%{\bf (especially because the 1994 occultation requires a non homogeneous ring)}, 
the other lines of evidence presented in this paper and the similarities with Chariklo are sufficient to make a strong case for the existence of a ring system in Chiron. On the other hand, alternative scenarios such as cometary jets or a cometary shell for the 2011 occultation features cannot explain all the observations. The fact that there can be two ring systems in the two largest centaurs might be an indication that rings could be common in centaurs and perhaps even in small TNOs, especially if the color bi-modality has something to do with the presence of rings.
During the revision phase of this paper the referee pointed out that a paper on the 2011 stellar occultation was submitted by \cite{Ruprecht2015} and a draft was kindly provided to us by the authors. We refer the reader to that publication for their most updated information on that occultation. Their timing remains the same as in \cite{Ruprecht2013a}, therefore our calculations need no changes.

\section*{Acknowledgments}

We are grateful to the CAHA and CASLEO staffs. This research is partially based on observations collected at the Centro Astron\'{o}mico Hispano Alem\'{a}n (CAHA) at Calar Alto, operated jointly by the Max-Planck Institut fur Astronomie and the Instituto de Astrof\'{i}sica de Andaluc\'{i}a (CSIC). This research was also partially based on observations carried out at the Complejo Astron\'{o}mico El Leoncito, operated under agreement between the Consejo Nacional
de Investigaciones Cient\'{i}ficas y T\'{e}cnicas de la Rep\'{u}blica Argentina and
the National Universities of La Plata, C\'{o}rdoba, and San Juan. Funding from Spanish grant AYA-2011-30106-CO2-O1 is acknowledged, as well as the Proyecto de Excelencia de la Junta de Andaluc\'{i}a, J.A. 2012-FQM1776 and FEDER funds. RD acknowledges the support of MINECO for his Ramon y Cajal Contract. AAC thanks FAPERJ and CNPq for financial support through diverse grants.  We are grateful to an anonymous referee for helping us to improve the paper.
%The CCD and data acquisition system at CASLEO were financed by R. M. Rich through U.S. NSF grant AST-90-15827.
  \\

%
%________________________________________________________________

%\begin{thebibliography}

\bibliographystyle{aa}
\bibliography{BiblioChiron_noteadded}

%\end{thebibliography}

%-------------------------------------------------------------------
\onecolumn

\begin{center}
%\begin{longtab}
\begin{longtable}{c c c c c c}
\caption{\label{tab:photo} Photometry results for the CAHA and CASLEO observations, respectively. In this sample table we list the Julian date (JD, corrected for light time), the relative magnitude (Rel. Mag in magnitudes) and the 1-$\sigma$ error associated (Err. in magnitude), the topocentric (r$_{h}$) and heliocentric ($\Delta$) distances (both distances expressed in AU) and the solar phase angle ($\alpha$, in degree). The full table is available in online.  } \\
\hline\hline
\multicolumn{6}{c}{CAHA 1.23m observations}\\
\hline\hline
 JD  & Rel. Mag.   & Err.  &   $\Delta$ & r$_\mathrm{h}$ & $\alpha$        \\
      & [mag]          &  [mag]&   [AU]      &  [AU]                   & [$^{\circ}$]    \\
\hline
\endfirsthead
\caption{continued.}\\
\hline\hline
 JD  & Rel. mag.   & Err.  &   $\Delta$ & r$_\mathrm{h}$ & $\alpha$        \\
      & [mag]          &  [mag]&   [AU]      &  [AU]                   & [$^{\circ}$]    \\
\hline
\endhead
\hline
\endfoot
2451783.41511  &   0.011  &   0.027  &   10.050   &  10.227   &  5.622  \\

2451783.41679  &   0.014  &   0.029  &   10.051   &  10.227   &  5.622  \\

2451783.41850  &   0.003  &   0.023  &   10.051   &  10.227   &  5.622  \\

2451783.42010  &   0.012  &   0.028  &   10.051   &  10.227   &  5.622  \\

2451783.42209  &   0.007  &   0.014  &   10.051   &  10.227   &  5.622  \\

2451783.42362  &   0.009  &   0.022  &   10.051   &  10.227   &  5.622  \\

2451783.42541  &   0.003  &   0.027  &   10.051   &  10.227   &  5.622  \\

2451783.42690  &  -0.009  &   0.021  &   10.051   &  10.227   &  5.622  \\

2451783.42847  &  -0.005  &   0.029  &   10.051   &  10.227   &  5.622  \\

2451783.43001  &  -0.043  &   0.041  &   10.051   &  10.227   &  5.622  \\

2451783.43183  &  -0.014  &   0.030  &   10.051   &  10.227   &  5.622  \\

2451783.43370  &  -0.025  &   0.032  &   10.051   &  10.227   &  5.622  \\

2451783.44383  &  -0.064  &   0.038  &   10.051   &  10.227   &  5.622  \\

2451783.44574  &  -0.066  &   0.040  &   10.051   &  10.227   &  5.622  \\

2451783.44724  &  -0.068  &   0.022  &   10.051   &  10.227   &  5.622  \\

2451783.44906  &  -0.078  &   0.011  &   10.051   &  10.227   &  5.622  \\

2451783.45056  &  -0.054  &   0.023  &   10.051   &  10.227   &  5.622  \\

2451783.45204  &  -0.062  &   0.030  &   10.051   &  10.227   &  5.622  \\

2451783.45350  &  -0.057  &   0.032  &   10.051   &  10.227   &  5.622  \\       

2451783.45510  &  -0.055  &   0.023  &   10.051   &  10.227   &  5.623  \\

2451783.45658  &  -0.060  &   0.018  &   10.051   &  10.227   &  5.623  \\

2451783.45810  &  -0.049  &   0.026  &   10.051   &  10.227   &  5.623  \\

2451783.46005  &  -0.050  &   0.022  &   10.051   &  10.227   &  5.623  \\

2451783.47206  &  -0.063  &   0.038  &   10.051   &  10.227   &  5.623  \\

2451783.47387  &  -0.040  &   0.037  &   10.052   &  10.227   &  5.623  \\

2451783.47554  &  -0.045  &   0.014  &   10.052   &  10.227   &  5.623  \\

2451783.47713  &  -0.050  &   0.026  &   10.052   &  10.227   &  5.623  \\

2451783.47864  &  -0.059  &   0.019  &   10.052   &  10.227   &  5.623  \\

2451783.48030  &  -0.032  &   0.036  &   10.052   &  10.227   &  5.623  \\

2451783.48178  &  -0.042  &   0.017  &   10.052   &  10.227   &  5.623  \\

2451783.48410  &  -0.046  &   0.027  &   10.052   &  10.227   &  5.623  \\

2451783.48616  &  -0.032  &   0.034  &   10.052   &  10.227   &  5.623  \\

2451783.49638  &  -0.021  &   0.023  &   10.052   &  10.227   &  5.623  \\

2451783.49810  &  -0.021  &   0.033  &   10.052   &  10.227   &  5.623  \\

2451783.49980  &  -0.005  &   0.022  &   10.052   &  10.227   &  5.623  \\

2451783.50159  &   0.005  &   0.015  &   10.052   &  10.227   &  5.623  \\

2451783.50332  &   0.004  &   0.022  &   10.052   &  10.227   &  5.623  \\

2451783.50509  &   0.005  &   0.018  &   10.052   &  10.227   &  5.623  \\

2451783.50707  &  -0.007  &   0.017  &   10.052   &  10.227   &  5.623  \\

2451783.50887  &   0.005  &   0.028  &   10.052   &  10.227   &  5.623  \\

2451783.51076  &   0.027  &   0.029  &   10.052   &  10.227   &  5.623  \\

2451783.51246  &   0.019  &   0.007  &   10.052   &  10.227   &  5.623  \\

2451785.42925  &  -0.079  &   0.051  &   10.087   &  10.230   &  5.638  \\

2451785.43285  &  -0.073  &   0.051  &   10.087   &  10.230   &  5.638  \\

2451785.43490  &  -0.070  &   0.055  &   10.087   &  10.230   &  5.638  \\

2451785.43668  &  -0.058  &   0.056  &   10.087   &  10.230   &  5.638  \\

2451785.43851  &  -0.041  &   0.056  &   10.087   &  10.230   &  5.638  \\

2451785.44105  &  -0.042  &   0.048  &   10.087   &  10.230   &  5.638  \\

2451785.44262  &  -0.047  &   0.055  &   10.087   &  10.230   &  5.638  \\

2451785.44513  &  -0.044  &   0.055  &   10.087   &  10.230   &  5.638  \\

2451785.44732  &  -0.027  &   0.053  &   10.087   &  10.230   &  5.638  \\

2451785.44940  &  -0.034  &   0.051  &   10.087   &  10.230   &  5.638  \\

2451785.45286  &  -0.012  &   0.051  &   10.087   &  10.230   &  5.639  \\

2451785.45461  &  -0.001  &   0.053  &   10.087   &  10.230   &  5.639  \\

2451785.45631  &  -0.020  &   0.054  &   10.087   &  10.230   &  5.639  \\

2451785.45848  &   0.009  &   0.055  &   10.087   &  10.230   &  5.639  \\

2451785.46031  &  -0.000  &   0.051  &   10.087   &  10.230   &  5.639  \\

2451785.46408  &   0.002  &   0.055  &   10.088   &  10.230   &  5.639  \\

2451785.46648  &   0.019  &   0.055  &   10.088   &  10.230   &  5.639  \\

2451785.46914  &   0.016  &   0.054  &   10.088   &  10.230   &  5.639  \\

2451785.47097  &   0.022  &   0.055  &   10.088   &  10.230   &  5.639  \\

2451785.47279  &   0.058  &   0.053  &   10.088   &  10.230   &  5.639  \\

2451785.47457  &   0.026  &   0.054  &   10.088   &  10.230   &  5.639  \\

2451785.47629  &   0.071  &   0.053  &   10.088   &  10.230   &  5.639  \\

2451785.47798  &   0.034  &   0.050  &   10.088   &  10.230   &  5.639  \\

2451785.47970  &   0.056  &   0.053  &   10.088   &  10.230   &  5.639  \\

2451785.48137  &   0.066  &   0.051  &   10.088   &  10.230   &  5.639  \\

2451785.48306  &   0.039  &   0.050  &   10.088   &  10.230   &  5.639  \\

2451785.48485  &   0.055  &   0.051  &   10.088   &  10.230   &  5.639  \\

2451785.48660  &   0.039  &   0.049  &   10.088   &  10.230   &  5.639  \\

2451785.48833  &   0.057  &   0.049  &   10.088   &  10.230   &  5.639  \\

2451785.49055  &   0.059  &   0.048  &   10.088   &  10.230   &  5.639  \\

2451785.49262  &   0.045  &   0.050  &   10.088   &  10.230   &  5.639  \\

2451785.49457  &   0.048  &   0.051  &   10.088   &  10.230   &  5.639  \\

2451785.49679  &   0.047  &   0.051  &   10.088   &  10.230   &  5.639  \\

2451785.50033  &   0.020  &   0.052  &   10.088   &  10.230   &  5.639  \\

2451785.50291  &   0.032  &   0.052  &   10.088   &  10.230   &  5.639  \\

2451785.50488  &   0.022  &   0.051  &   10.088   &  10.230   &  5.639  \\

2451785.50694  &   0.015  &   0.052  &   10.088   &  10.230   &  5.639  \\

2451785.50876  &   0.040  &   0.052  &   10.088   &  10.231   &  5.639  \\

2451785.51043  &   0.012  &   0.051  &   10.088   &  10.231   &  5.639  \\

2451785.51282  &   0.037  &   0.052  &   10.088   &  10.231   &  5.639  \\

2451785.51460  &   0.043  &   0.051  &   10.088   &  10.231   &  5.639  \\

2451785.51675  &   0.032  &   0.051  &   10.089   &  10.231   &  5.639  \\

2451785.51843  &   0.028  &   0.051  &   10.089   &  10.231   &  5.639  \\

2451785.52374  &   0.022  &   0.049  &   10.089   &  10.231   &  5.639  \\

2451785.52591  &   0.012  &   0.051  &   10.089   &  10.231   &  5.639  \\

2456575.26643  &  -0.013  &   0.015  &   16.836   &  17.643   &  1.951  \\

2456575.26945  &   0.006  &   0.015  &   16.836   &  17.643   &  1.951  \\

2456575.27248  &  -0.001  &   0.015  &   16.836   &  17.643   &  1.951  \\

2456575.27551  &  -0.007  &   0.015  &   16.836   &  17.643   &  1.951  \\

2456575.27853  &  -0.006  &   0.015  &   16.837   &  17.643   &  1.951  \\

2456575.28155  &  -0.017  &   0.015  &   16.837   &  17.643   &  1.951  \\

2456575.28457  &  -0.004  &   0.015  &   16.837   &  17.643   &  1.951  \\

2456575.28759  &  -0.022  &   0.015  &   16.837   &  17.643   &  1.951  \\

2456575.29061  &   0.009  &   0.015  &   16.837   &  17.643   &  1.952  \\

2456575.29364  &  -0.000  &   0.015  &   16.837   &  17.643   &  1.952  \\

2456575.29667  &   0.004  &   0.015  &   16.837   &  17.643   &  1.952  \\

2456575.29969  &  -0.017  &   0.015  &   16.837   &  17.643   &  1.952  \\

2456575.30271  &   0.023  &   0.015  &   16.837   &  17.643   &  1.952  \\

2456575.30573  &   0.014  &   0.015  &   16.837   &  17.643   &  1.952  \\

2456575.30875  &   0.000  &   0.015  &   16.837   &  17.643   &  1.952  \\

2456575.31178  &   0.003  &   0.015  &   16.837   &  17.643   &  1.953  \\

2456575.31480  &   0.008  &   0.015  &   16.837   &  17.643   &  1.953  \\

2456575.31782  &   0.011  &   0.015  &   16.837   &  17.643   &  1.953  \\

2456575.32084  &   0.016  &   0.015  &   16.837   &  17.643   &  1.953  \\

2456575.32442  &   0.002  &   0.015  &   16.837   &  17.643   &  1.953  \\

2456575.32686  &  -0.003  &   0.015  &   16.837   &  17.643   &  1.953  \\ 

2456575.32932  &   0.040  &   0.014  &   16.837   &  17.643   &  1.953  \\

2456575.33176  &  -0.006  &   0.015  &   16.837   &  17.643   &  1.953  \\

2456575.33420  &   0.004  &   0.015  &   16.837   &  17.643   &  1.954  \\

2456575.33664  &  -0.003  &   0.015  &   16.837   &  17.643   &  1.954  \\

2456575.33908  &   0.010  &   0.015  &   16.837   &  17.643   &  1.954  \\

2456575.34153  &  -0.004  &   0.015  &   16.837   &  17.643   &  1.954  \\

2456575.34398  &   0.006  &   0.015  &   16.837   &  17.643   &  1.954  \\

2456575.34642  &   0.010  &   0.015  &   16.837   &  17.643   &  1.954  \\

2456575.34886  &  -0.003  &   0.015  &   16.837   &  17.643   &  1.954  \\

2456575.35131  &  -0.002  &   0.015  &   16.837   &  17.643   &  1.954  \\

2456575.35375  &   0.001  &   0.015  &   16.837   &  17.643   &  1.954  \\ 

2456575.35619  &   0.001  &   0.015  &   16.837   &  17.643   &  1.955  \\

2456575.35864  &   0.010  &   0.015  &   16.837   &  17.643   &  1.955  \\

2456575.36109  &   0.019  &   0.015  &   16.837   &  17.643   &  1.955  \\ 

2456575.36352  &  -0.002  &   0.015  &   16.837   &  17.643   &  1.955  \\

2456575.36596  &  -0.004  &   0.015  &   16.837   &  17.643   &  1.955  \\

2456575.36840  &   0.005  &   0.015  &   16.838   &  17.643   &  1.955  \\

2456575.37085  &  -0.025  &   0.015  &   16.838   &  17.643   &  1.955  \\

2456575.37330  &   0.014  &   0.015  &   16.838   &  17.643   &  1.955  \\

2456575.37574  &   0.001  &   0.015  &   16.838   &  17.643   &  1.955  \\

2456575.37818  &   0.005  &   0.015  &   16.838   &  17.643   &  1.955  \\

2456575.38063  &   0.014  &   0.015  &   16.838   &  17.643   &  1.955  \\

2456575.38728  &  -0.007  &   0.015  &   16.838   &  17.644   &  1.956  \\

2456575.38972  &  -0.006  &   0.015  &   16.838   &  17.644   &  1.956  \\

2456575.39216  &  -0.002  &   0.015  &   16.838   &  17.644   &  1.956  \\

2456575.39461  &  -0.023  &   0.015  &   16.838   &  17.644   &  1.956  \\

2456575.39705  &   0.000  &   0.015  &   16.838   &  17.644   &  1.956  \\

2456575.39949  &  -0.003  &   0.015  &   16.838   &  17.644   &  1.957  \\

2456575.40194  &   0.007  &   0.015  &   16.838   &  17.644   &  1.957  \\

2456575.40439  &  -0.012  &   0.015  &   16.838   &  17.644   &  1.957  \\ 

2456575.40683  &   0.017  &   0.015  &   16.838   &  17.644   &  1.957  \\

2456575.40927  &  -0.002  &   0.015  &   16.838   &  17.644   &  1.957  \\

2456575.41171  &   0.003  &   0.015  &   16.838   &  17.644   &  1.957  \\

2456575.41416  &  -0.001  &   0.015  &   16.838   &  17.644   &  1.957  \\

2456575.41661  &   0.009  &   0.015  &   16.838   &  17.644   &  1.957  \\

2456575.41905  &   0.010  &   0.015  &   16.838   &  17.644   &  1.957  \\

2456575.42149  &   0.004  &   0.015  &   16.838   &  17.644   &  1.958  \\

2456575.42394  &   0.016  &   0.015  &   16.838   &  17.644   &  1.958  \\

2456575.42638  &   0.002  &   0.015  &   16.838   &  17.644   &  1.958  \\

2456575.42882  &  -0.011  &   0.015  &   16.838   &  17.644   &  1.958  \\

2456575.43127  &   0.028  &   0.014  &   16.838   &  17.644   &  1.958  \\

2456575.43372  &  -0.012  &   0.015  &   16.838   &  17.644   &  1.958  \\

2456575.43616  &   0.001  &   0.015  &   16.838   &  17.644   &  1.958  \\

2456575.43860  &  -0.003  &   0.015  &   16.838   &  17.644   &  1.958  \\ 

2456575.44104  &  -0.008  &   0.015  &   16.838   &  17.644   &  1.958  \\

2456575.44350  &  -0.007  &   0.015  &   16.838   &  17.644   &  1.959  \\

2456575.44594  &   0.004  &   0.015  &   16.838   &  17.644   &  1.959  \\

2456576.18288  &   0.011  &   0.015  &   16.847   &  17.644   &  1.991  \\

2456576.18893  &   0.030  &   0.014  &   16.847   &  17.644   &  1.992  \\

2456576.19195  &   0.014  &   0.015  &   16.847   &  17.644   &  1.992  \\

2456576.19497  &  -0.009  &   0.015  &   16.847   &  17.644   &  1.992  \\

2456576.19799  &   0.008  &   0.015  &   16.847   &  17.644   &  1.992  \\

2456576.20102  &   0.018  &   0.015  &   16.847   &  17.644   &  1.992  \\

2456576.20636  &  -0.009  &   0.015  &   16.847   &  17.644   &  1.993  \\

2456576.20938  &  -0.009  &   0.015  &   16.847   &  17.644   &  1.993  \\

2456576.21240  &  -0.023  &   0.015  &   16.847   &  17.644   &  1.993  \\

2456576.21542  &  -0.021  &   0.015  &   16.847   &  17.644   &  1.993  \\ 

2456576.21846  &  -0.016  &   0.015  &   16.847   &  17.644   &  1.993  \\

2456576.22148  &   0.008  &   0.015  &   16.847   &  17.644   &  1.993  \\

2456576.22450  &  -0.025  &   0.015  &   16.847   &  17.644   &  1.993  \\

2456576.23054  &  -0.014  &   0.015  &   16.847   &  17.644   &  1.994  \\

2456576.23356  &  -0.026  &   0.015  &   16.847   &  17.644   &  1.994  \\

2456576.23961  &  -0.040  &   0.015  &   16.847   &  17.644   &  1.994  \\

2456576.26983  &  -0.024  &   0.015  &   16.848   &  17.644   &  1.995  \\

2456576.27286  &  -0.032  &   0.015  &   16.848   &  17.644   &  1.995  \\

2456576.27890  &  -0.014  &   0.015  &   16.848   &  17.644   &  1.996  \\

2456576.28192  &  -0.006  &   0.015  &   16.848   &  17.644   &  1.996  \\

2456576.28494  &  -0.007  &   0.015  &   16.848   &  17.644   &  1.996  \\

2456576.28797  &   0.001  &   0.015  &   16.848   &  17.644   &  1.996  \\

2456576.29517  &  -0.004  &   0.015  &   16.848   &  17.644   &  1.996  \\

2456576.29819  &   0.007  &   0.015  &   16.848   &  17.644   &  1.997  \\

2456576.30121  &   0.022  &   0.015  &   16.848   &  17.644   &  1.997  \\

2456576.30423  &   0.007  &   0.015  &   16.848   &  17.644   &  1.997  \\

2456576.30725  &   0.013  &   0.015  &   16.848   &  17.644   &  1.997  \\

2456576.31028  &   0.013  &   0.015  &   16.848   &  17.644   &  1.997  \\

2456576.31331  &   0.014  &   0.015  &   16.848   &  17.644   &  1.997  \\

2456576.31633  &   0.016  &   0.015  &   16.848   &  17.644   &  1.997  \\

2456576.31935  &   0.004  &   0.015  &   16.848   &  17.644   &  1.998  \\

2456576.32237  &   0.014  &   0.015  &   16.848   &  17.644   &  1.998  \\

2456576.32539  &  -0.012  &   0.015  &   16.848   &  17.644   &  1.998  \\

2456576.32842  &   0.016  &   0.015  &   16.848   &  17.644   &  1.998  \\

2456576.33144  &   0.014  &   0.015  &   16.848   &  17.644   &  1.998  \\

2456576.33447  &   0.013  &   0.015  &   16.848   &  17.644   &  1.998  \\

2456576.33749  &   0.012  &   0.015  &   16.848   &  17.644   &  1.998  \\

2456576.34051  &   0.011  &   0.015  &   16.848   &  17.644   &  1.998  \\ 

2456576.34353  &   0.006  &   0.015  &   16.849   &  17.644   &  1.999  \\

2456576.34656  &  -0.006  &   0.015  &   16.849   &  17.644   &  1.999  \\

2456576.34958  &   0.013  &   0.015  &   16.849   &  17.644   &  1.999  \\

2456576.35260  &   0.000  &   0.015  &   16.849   &  17.644   &  1.999  \\

2456576.35562  &  -0.006  &   0.015  &   16.849   &  17.644   &  1.999  \\

2456576.35864  &  -0.008  &   0.015  &   16.849   &  17.644   &  1.999  \\ 

2456576.36166  &   0.007  &   0.015  &   16.849   &  17.644   &  1.999  \\

2456576.36470  &  -0.008  &   0.015  &   16.849   &  17.644   &  1.100  \\

2456576.36772  &   0.017  &   0.015  &   16.849   &  17.644   &  1.100  \\

2456576.37074  &   0.009  &   0.015  &   16.849   &  17.644   &  1.100  \\

2456576.37376  &   0.015  &   0.015  &   16.849   &  17.644   &  2.000  \\

2456576.37678  &   0.012  &   0.015  &   16.849   &  17.644   &  2.000  \\

2456576.37980  &   0.014  &   0.015  &   16.849   &  17.644   &  2.000  \\

2456576.38283  &   0.020  &   0.015  &   16.849   &  17.644   &  2.001  \\

2456576.38585  &   0.026  &   0.014  &   16.849   &  17.644   &  2.001  \\

2456576.38888  &   0.007  &   0.015  &   16.849   &  17.644   &  2.001  \\

2456576.39190  &   0.007  &   0.015  &   16.849   &  17.644   &  2.001  \\

2456576.39492  &   0.007  &   0.015  &   16.849   &  17.644   &  2.001  \\

2456576.39795  &   0.019  &   0.015  &   16.849   &  17.644   &  2.001  \\

2456576.40097  &   0.004  &   0.015  &   16.849   &  17.644   &  2.001  \\

2456576.40734  &  -0.010  &   0.015  &   16.849   &  17.644   &  2.002  \\

2456576.41036  &   0.010  &   0.015  &   16.849   &  17.644   &  2.002  \\

2456576.41339  &  -0.019  &   0.015  &   16.849   &  17.644   &  2.002  \\

2456576.41641  &  -0.014  &   0.015  &   16.849   &  17.644   &  2.002  \\

2456576.41943  &  -0.027  &   0.015  &   16.850   &  17.644   &  2.002  \\

2456576.42244  &  -0.029  &   0.015  &   16.850   &  17.644   &  2.002  \\

2456576.42546  &  -0.033  &   0.015  &   16.850   &  17.644   &  2.002  \\

2456576.42848  &  -0.033  &   0.015  &   16.850   &  17.644   &  2.002  \\

2456576.43756  &  -0.037  &   0.015  &   16.850   &  17.644   &  2.003  \\

2456576.44360  &  -0.027  &   0.015  &   16.850   &  17.644   &  2.003  \\

\hline
\hline
\end{longtable}
%\end{longtab}
\end{center}

\end{document}